%% file: RR-9395.tex
\documentclass[twoside]{article}
\usepackage[a4paper]{geometry}

\usepackage[utf8]{inputenc}

\usepackage{RR}
\usepackage{graphicx}
\usepackage[table]{xcolor}
\usepackage{floatrow}
\usepackage[obeyspaces]{url} 
\usepackage[numbers,square,sort&compress]{natbib} 
\usepackage{amsmath}
\usepackage{amssymb}
\usepackage{slashbox}
\usepackage{siunitx}
\usepackage{booktabs}
\usepackage{multirow}
\usepackage{pifont}
\usepackage[defblank]{paralist}  
\usepackage[inline]{enumitem}
\usepackage{subfig}

\usepackage{skull}
\usepackage{./why3lang}
\usepackage{tikz}
\usetikzlibrary{shapes.geometric, arrows, patterns}
\usetikzlibrary{backgrounds,calc,trees}
\usepackage{stmaryrd}
\usepackage{mathpartir}
\usepackage{ifthen}
\newboolean{showcomments}
\setboolean{showcomments}{false} 

\ifthenelse{\boolean{showcomments}}
{\newcommand{\nbnote}[3]{
  \fcolorbox{gray}{yellow}{\bfseries\sffamily\scriptsize#1}
  {\color{#2} \sffamily\small$\blacktriangleright$\textit{#3}$\blacktriangleleft$}
  }
}
{\newcommand{\nbnote}[3]{}
 
}

\usepackage{xcolor}
\usepackage{colortbl}
\usepackage{rotating}

\usepackage[hidelinks]{hyperref}

\usepackage[nameinlink]{cleveref}
\crefname{subsection}{subsection}{subsections}

\newcommand{\provername}[1]{\cellcolor{yellow!25}
\begin{sideways}\textbf{#1}~~\end{sideways}}
\newcommand{\explanation}[1]{\cellcolor{yellow!13}lemma \texttt{#1}}
\newcommand{\continuation}[1]{\cellcolor{yellow!13}\texttt{#1}}

\newcommand{\valid}[1]{\cellcolor{green!13}#1}

\newcommand{\timeout}[1]{\cellcolor{red!20}(#1)}

\newcommand{\noresult}{\multicolumn{1}{>{\columncolor[gray]{0.8}}c|}{~}}

\newcommand{\highfailure}{\footnotesize{\cellcolor{red!50}FAILURE}}
\newcommand{\seqnodes}{\mathit{seq}(\!\Nodes\!)}

\pgfdeclarelayer{background}
\pgfsetlayers{background,main}

\newcommand{\Inv}{\mathit{Inv}}
\newcommand{\Pre}{\mathit{Pre}}
\newcommand{\Post}{\mathit{Post}}
\newcommand{\Nodes}{\mathit{Nodes}}
\newcommand{\troot}{\mathit{root}}
\newcommand{\TS}{\mathit{TS}}
\newcommand{\rank}{\mathit{rank}}

\newcommand{\RDTName}{{Maram}}

\newcommand{\callout}[2][yellow]{\mathchoice%
	{\colorbox{#1}{$\displaystyle#2$}}%
	{\colorbox{#1}{$\textstyle#2$}}%
	{\colorbox{#1}{$\scriptstyle#2$}}%
	{\colorbox{#1}{$\scriptscriptstyle#2$}}}%

\definecolor{darkgreen}{rgb}{0.0, 0.5, 0.0}

\newcommand\focus[1]{\callout[gray!30]{\begin{color}{blue}#1\end{color}}}

\newcommand\echanged[1]{\callout[blue!20]{#1}}
\newcommand\enew[1]{\callout[red!20]{#1}}

\newcommand\tabfocus[1]{\begin{color}{blue}#1\end{color}}


\mathchardef\mhyphen="2D
\RRetitle{A coordination-free, convergent, and safe replicated tree}

\RRdate{January 2022}
 
  \RRtitle{Un arbre r\'epliqu\'e, convergent et s\^ur sans coordination}
    \RRresume{  L'arbre est une structure de donn\'ees essentielle.  Quand l'application est distribu\'ee, par exemple dans un syst\`eme de fichiers distribu\'e, l'arbre est r\'epliqu\'e.
  Pour am\'eliorer les performances et la disponibilit\'e, les diff\'erents clients doivent pouvoir mettre \`a jour leurs r\'epliques simultan\'ement et sans coordination.  Celles-ci convergent si les mises \`a jour commutent entre elles ; n\'eanmoins, m\^eme dans ce cas, des op\'erations ``move'' concurrentes peuvent conduire \`a des \'etats incorrects, et m\^eme \`a la perte de donn\'ees.  Au bout du compte, entre deux op\'erations ``move'' en conflit, seul  l'une des deux peut \^etre autoris\'ee \`a prendre effet. 
  Cependant, comme ce cas est rare, la solution doit \^etre l\'eg\`ere.
  Les approches pr\'ec\'edentes n\'ecessitaient une coordination pr\'eventive des r\'epliques, ou des retours en arri\`ere \`a posteriori.

  Dans cet article, nous pr\'esentons un nouvel arbre r\'epliqu\'e, qui met en {\oe}uvre une op\'eration ``move'' atomique sans coordination, et dont nous prouvons qu'il maintient l'invariant d'arbre.
  Notre analyse identifie les cas o\'u les ``move'' concurrents sont intrins\`equement s\^urs, et proposons un algorithme l\'eger, sans coordination et sans retour-arri\`ere, pour les autres cas, de sorte qu'un sous-ensemble maximal et s\^ur de ``move'' prenne effet.

  Nous pr\'esentons une analyse d\'etaill\'ee des probl\`emes de coh\'erence dans les arbres.
  Nous fournissons une preuve m\'ecanis\'ee que la structure des donn\'ees est convergente et maintient l'invariant d'arbre.
  Enfin, nous comparons le temps de r\'eponse et la disponibilit\'e de notre concept
  \`a la litt\'erature.  }
 
 \RRabstract{
  The tree is an essential data structure in many applications.
  In a distributed application, such as a distributed file system, the tree is replicated.
  To improve performance and availability, different clients should be able to update their replicas concurrently and without coordination.
  Such concurrent updates converge if the effects commute, but nonetheless, concurrent moves can lead to incorrect states and even data loss.
  Such a severe issue cannot be ignored; ultimately, only one of the conflicting moves may be allowed to take effect.
  However, as it is rare, a solution should be lightweight.
  Previous approaches would require preventative cross-replica coordination, or totally order all operations after-the-fact, requiring roll-back and compensation operations.

  In this paper, we present a novel replicated tree that
  supports coordination-free concurrent atomic moves, and provably
  maintains the tree invariant.
  Our analysis identifies cases where concurrent moves are inherently
  safe, and we devise a lightweight, coordination-free, rollback-free algorithm for the
  remaining cases, such that a maximal safe subset of moves takes effect.

  We present a detailed analysis of the concurrency issues with trees, justifying our replicated tree data structure.
  We provide mechanized proof that the data structure is convergent and maintains the tree invariant.
  Finally, we compare the response time and availability of our design
  against the literature.

  }
\RRauthor{Sreeja S. Nair\thanks{Sorbonne Universit\'e---LIP6 \& Inria}, Filipe Meirim\thanks{NOVA LINCS, Universidade Nova de Lisboa}, M\'ario Pereira\thanks{NOVA LINCS, Universidade Nova de Lisboa}, Carla Ferreira\thanks{NOVA LINCS, Universidade Nova de Lisboa}, Marc Shapiro\thanks{Sorbonne Universit\'e---LIP6 \& Inria}}
     \RRkeyword{Distributed data structures, Conflict-free Replicated Data Type, Formal verification}
    \RRmotcle{Structures de donn\'ees distribu\'ees, CRDT, V\'erification formelle}

\RRprojet{DELYS}
    \RCParis

\begin{document}
\RRNo{9395}
\makeRR

\input{sections/intro.tex}

\input{sections/system.tex}
\input{sections/incorrectspec.tex}

\input{sections/refinedspec.tex}
\input{sections/evaluation.tex}
\input{sections/literature.tex}
\input{sections/discussion.tex}

\input{sections/conclusion.tex}

\bibliographystyle{plainnat}
\bibliography{predef,references}

\newpage

\appendix
\input{sections/why3spec.tex}

\end{document}

%% file: sections/intro.tex
\section{Introduction}

Concurrent data structures are an important programming abstraction; 
designing concurrent data structures with non-trivial properties is
complex.
The tree data structure is used in many applications.
For instance, a file system is a tree of directories and files.
A move (or rename) operation transfers a subtree atomically by changing its parent.
Similarly, a rich text editor maintains a DOM tree of blocks with attributes.
Text editing modifies the tree structure; in particular a \emph{drag and drop} can move a subtree from one parent to another.

A tree has a 
strong structural invariant: nodes are unique, there is
a single root, each node has a single parent and has a path to the root,
and the child-parent graph is acyclic. Much current work in concurrent data structure design focuses on
lock-free or wait-free coordination using primitives such as
compare-and-swap (CAS).
However, in a distributed and replicated setting, even CAS is too
strong.
Consider a file system replicated to several locations over
the globe, or through a mobile network.
Network latency between continents can be between 0.1 and 0.5
seconds; the mobile network may disconnect completely.
To ensure availability, a user of the file system must be able to update
a replica locally, and update \emph{without coordinating at all} with
other replicas.
Replicas converge eventually by exchanging their updates asynchronously.


It is a major challenge to maintain safety in this context;
specifically, in this case, to maintain the tree structure. Concurrent atomic moves (also called renames in a file system) are especially problematic \cite{syn:rep:1652}.
Consider for instance a tree composed of the root and children $a$ and
$b$.
One replica moves $a$ underneath $b$, while concurrently (without
coordination) the other replica moves $b$ under $a$.
Na{\" \i}vely replaying one replica's updates at the other
produces an $a-b$ cycle disconnected from the root.

This is a widespread issue; indeed, many replicated file systems have
serious anomalies, including incorrect or diverged states \cite[Section 6 for some examples]{fic:rep:sh172}, violating the tree
invariant~\cite{syn:rep:1652}.
However, concurrent moves are relatively rare in these systems\footnote{For example, a file system trace we analyzed contained 1198823 operations in total, 20883 create operations, 49509 remove operations and just 547 move operations (70939 structural operations altogether).}
and it is important that we design a solution that has minimal overhead. Solutions in the literature include non-atomic moves~\cite{fic:rep:sh172} (resulting in duplicate copies),
re-introducing coordination~\cite{formel:rep:syn:sh197} (first one to acquire lock proceeds; others abort), or requiring
roll-backs~\cite{DBLP:journals/corr/abs-1805-04263} (the move operation ordered first proceeds,  all concurrent operations are rolled back).
\citet{formel:rep:syn:sh197} shows that there can be no \mbox{coordination-free solution to this problem that is not  anomalous.}

To support low latency, high availability,  and safety,
this paper introduces a new light-weight, coordination-free, safe, replicated
CRDT~\cite{syn:rep:sh143} tree data structure, called
\emph{\RDTName}.
\RDTName\ supports the usual operations to query the state, to add or to
remove a node, and also supports an atomic~\emph{move} operation.
The price to pay is that some {move} operations ``lose'', i.e., have
no effect; achieving the same end result as previous correct approaches but at a lower cost. Query and add are unremarkable. 
Remove marks the corresponding node as a ``tombstone,'' but leaves it in
the data structure, as is common in replicated data structures~\cite{rep:syn:app:1765}.
We show that moves can be divided into two cases: two concurrent
\emph{up-move}s are always safe.
We devise a deterministic arbitration rule for conflicts of
\emph{down-move}: against a concurrent up-move, the up-move wins, and
the down-move loses; against a concurrent down-move, 
the down-move with the highest
priority (as defined in Section~\ref{sec:move-move}) wins and the other loses.

We prove \RDTName\ to be safe, even in the presence of concurrent
updates (including moves), despite being coordination-free and without
any roll-backs.
Using the Why3 proof assistant, we apply the CISE proof methodology~\cite{syn:app:sh179}, with the following steps:
\begin{compactenum}
\item
  \emph{Sequential safety:} We show that the initial state satisfies the
  tree invariant, and that every individual update operation has a
  precondition strong enough to maintain the  invariant.
\item
  \emph{Convergence:} We show that any two operations that may execute
  concurrently commute.
\item
  \emph{Precondition stability:} We show that for any two operations $u, v$
  that may execute concurrently, $u$ preserves the precondition of $v$,
  and vice-versa.
\end{compactenum}
It follows that every state reachable from the initial state, 
sequentially or concurrently, satisfies the tree invariant.%
\footnote{%
We furthermore claim ({without proof}) that \RDTName\ is live, in
  the sense that, if every message sent is eventually delivered to some
  replica $r_1$, then, given some update originating at a replica $r_2$, its
  postcondition eventually takes effect at replica $r_1$.
}
\RDTName{} satisfies an additional desirable property, \emph{monotonic
  reads}~\cite{rep:syn:1481}.
This requires that a replica that has delivered some update will not
roll it back.

Losing an operation might have impacts on the causally dependent future operations. 
We devise a independence analysis to capture this effect. 
The move operations lose if it is dependent on another move operation that loses.

This paper presents the principles of \RDTName{}, proves its
correctness, and compares the performance of \RDTName\ to competing
solutions in a simulated geo-replicated environment.
The response time of \RDTName\ is 1.35 times of the safe rollback-based design, and 1.36 times of the unsafe uncoordinated design (both due to overhead of computing the metadata required for conflict resolution), and up to 11 times faster than (safe) lock-based designs.
Furthermore, \RDTName\ stabilises (updates become definitive) 
three orders of magnitude faster
than a safe rollback-based design.

This paper proceeds as follows.
\begin{inparablank}
\item
  Section~\ref{sec:backgd} formalises our system model, explains our
  proof methodology, and defines the tree invariant.
\item
  In Section~\ref{sec:seqdesign} we discuss the sequential correctness of
  a replicated tree.
\item
  Section~\ref{sec:refinedspec} proceeds with the proof of convergence,  precondition stability and independence, resulting in concurrent safety.
\item
  In Section~\ref{sec:evaluation} we compare the performance of \RDTName\
   with competing designs.
\item
  Section~\ref{sec:literature} overviews the related literature.
\item
  Finally, in Section~\ref{sec:discussion} we discuss lessons learned
  and their significance.  
\end{inparablank}

%% file: sections/system.tex

\section{Preliminaries}
\label{sec:backgd}


\subsection{System Model}
\label{sec:model}

A distributed system is modelled as a set of processes, distributed over
a (high-latency, failure-prone) communication network.
The processes have disjoint memory and processing capabilities, and they
communicate through message passing.
A process does not fail. 
Every message is eventually delivered to its destination.
Message delivery is consistent with happens-before (causal consistency).

\subsubsection*{State and invariant:}

The data structure (in this case, a tree) is
\emph{replicated} at a number of processes, called its \emph{replicas}.
The information managed by a replica on behalf of the data structure is
called its \emph{local state}.
The union of local states is called the
\emph{global state}.%
\footnote{
  {Note that this global view cannot be observed by any single replica
    and is merely an explanatory device.}
}

A data structure is associated with an \emph{invariant}, a predicate
that must always be satisfied in every local state of a replica.
Although evaluated locally, an invariant describes a global property, in
the sense that it must be true at all replicas.

\subsubsection*{Operations:}

An unspecified client application submits an operation at some replica
of its choice, which we call the \emph{origin} replica of that operation.
For availability, the origin replica should carry out the operation
without waiting to coordinate with other replicas.

An update operation has a
\emph{postcondition} that specifies the state after the operation
executes, and a 
\emph{precondition} that indicates the domain of the operation.
As discussed in more detail later, when the operation executes with no
concurrency, its precondition guarantees that the operation terminates
with the postcondition satisfied.

\subsubsection*{Updates:}
When a client submits an operation, the origin replica generates an
\emph{effector} (a side-effecting lambda), atomically applies the
effector to the origin state, and sends the effector to all the other
replicas.
Every replica eventually receives and delivers the effector, atomically
applying it to its own local state.%
\footnote{
  Note that, at this point, the system is committed to this operation, and
  the operation's precondition must be true at the remote replica.
}
The effector eventually executes at every replica.

We assume that effectors are delivered in causal order.
This means that, if some replica that observed an effector $u$ later
generates an effector $v$, then any replica that observes $v$ has
previously observed $u$.%
\footnote{
  In \Cref{sec:discussion} we consider relaxing this
  requirement to eventual consistency, which states only that all updates
  are eventually delivered at all replicas.
}

In what follows, we ignore queries, and identify an
update operation with executing its effector at all replicas.

\subsection{Properties and associated Proof Rules}
\label{sec:rules}

Consider some data structure (in this case a tree) characterized
by a safety \emph{invariant} (in this case, the tree invariant).
We say that a state is \emph{local-safe} if it satisfies the data
structure's invariant.
An update is \emph{op-safe} if, starting from a local-safe state, it
leaves it a local-safe state.
The distributed data structure is \emph{safe} if every update is
op-safe.
According to the CISE logic \cite{syn:app:sh179}, a distributed data
structure is safe if the following properties hold:
\begin{compactenum}
\item
  \emph{Sequential safety}: Consider an environment restricted to
  sequential execution (operations execute one after another; there is no concurrency).
  If the initial state is local-safe at every
  replica, and each update is op-safe, it follows that the data
  structure is safe under sequential execution.
  Classically, sequential op-safety implies that each operation's
  precondition satisfies the weakest-precondition of the invariant with
  respect to the operation \cite{formel:lan:1832}.
\item
  \emph{Convergence:} Strong Eventual Consistency (SEC)
  \cite{syn:rep:sh143} states that two replicas that have delivered the
  same set of operations must be in the same state, i.e., the system
  converges.
  If operations commute (as defined later), then SEC is guaranteed
  \cite{syn:rep:sh143}.
\item
  \emph{Precondition stability:} In addition to sequential safety,
  updates must remain op-safe in the presence of concurrent
  (uncoordinated) updates.
  To ensure this, we apply the CISE precondition stability rule
  \cite{syn:app:sh179}: consider two updates $u$ and $v$; if the execution
  of $u$ does not make the precondition of $v$ false, nor vice-versa
  (\emph{precondition stability}), then executing $u$ and $v$ concurrently
  is op-safe.
  This must be true for all concurrent pairs of operations.
\end{compactenum}

CISE logic helps us identify the conditions under which concurrent operations conflict.
When conflicting, CISE requires the operations to acquire tokens, that bring in a global synchronization point.
Hence all updates in CISE are assumed to be definitive.

In order to augment the CISE analysis for handling tentative updates, we add a condition for \emph{independence} to check whether skipping a move affects a move that already observed the effect of the skipped one.
The independency analysis is inspired from \citet{Houshmand:2019:HRC:3302515.3290387}, even though they also, like CISE, do not consider tentative updates.

\emph{Independence analysis:} Consider two updates $u$ and $v$ that are safe, $u$ executed before $v$. 
If moving $v$ before $u$ still maintains the safety of $v$, $v$ is said to be independent of $u$.
Otherwise, if $v$ is unsafe before executing $u$, $v$ is dependent on the effect of $u$.

\subsubsection{Sequential safety}
\label{sec:backgd:opsafe}
Let us refine the proof obligations of the first step, sequential
safety{}, i.e., local-safety under sequential execution.

The set of reachable states comprises the initial
state, and all states transitively reachable as a result of executing
updates sequentially.
The set of reachable states is a subset of the set of all possible states.
Formally, we note the set of states $\Sigma$, a state
$\sigma$, the initial state $\sigma_{init}$, an update $u$, its
precondition $\Pre_{u}$, and the set of updates $U$.
When execution is sequential:
\begin{small}
\begin{align}
\sigma_{\mathit{init}} \in \Sigma
\label{eq:1}
\end{align}
\end{small}
and 
\begin{small}
\begin{align}
  \forall u \in U , \sigma \in \Sigma  \centerdot \sigma \models \Pre_u \implies u(\sigma) \in \Sigma
  \label{eq:2}
\end{align}
\end{small}
$\Sigma$ is the smallest set satisfying (\ref{eq:1}) and (\ref{eq:2})
through a sequence of legal updates from the initial state.

The data structure must satisfy its invariant in every sequentially reachable state: this property is called \emph{sequential safety}.
Formally, if $\Inv$ denotes the invariant, then
\begin{small}
\begin{align}
  \forall \sigma \in \Sigma \centerdot \sigma \models \Inv \label{eq:safestates}
\end{align}
\end{small}

If the initial state is safe and all sequential updates preserve the invariant, by induction, the data structure is sequentially safe.
Formally, if the initial state, $\sigma_{init}$, satisfies the invariant, $\Inv$,
\begin{small}
\begin{align}
\sigma_{\mathit{init}} \models \Inv
\end{align}
\end{small}
and each update $u$ executing on a state $\sigma$ preserves the invariant,
\begin{small}
\begin{align}
\forall u \in U, \sigma, \sigma' \in \Sigma\, \centerdot\, \sigma \models (\Inv \wedge \Pre_u) \wedge u(\sigma) = \sigma' \implies \sigma' \models \Inv 
\label{seq-safety}
\end{align}
\end{small}
then the invariant holds true for all reachable states.
$Pre_u$ is the weakest precondition required to maintain the safety of update $u$. 
Weakest precondition for an update can be calculated by predicate transformer semantics as described by \citet{formel:lan:1832}.

\subsubsection{Concurrency}
\label{sec:rules:concurrency}
Let us now turn to concurrent execution, and consider the proof obligations for convergence and safety.

\paragraph{Convergence}
\label{sec:rules:concurrency:convergence}

If a replica initiates an update $u$, while concurrently another
replica initiates $v$, the first replica executes their effectors in the
order $u;v$ and the second one in the order $v;u$.
Without precaution, it is likely that their states diverge.

To prevent this, the Strong Eventual Consistency (SEC) property
\cite{syn:rep:sh143} requires that any two replicas that delivered the
same updates are in equivalent states.
To satisfy SEC, effector functions are designed to commute, i.e., both
orders above leave the data in the same state.
We define commutativity as follows:
\begin{small}
\begin{align}
\forall u_1, u_2 \in U, \sigma, \sigma_1, \sigma_2 \in \Sigma \centerdot u_1(\sigma) = \sigma_1  \wedge  u_2(\sigma)  = \sigma_2 \implies u_2(\sigma_1) = u_1(\sigma_2) 
\end{align}
\end{small}

\paragraph{Precondition stability} 
\label{sec:rules:concurrency:stability}
The main proof obligation, for concurrent execution, is that the
precondition of any effector is stable against (i.e., not negated by) an
effector that may execute concurrently \cite{syn:app:sh179}.
This CISE rule is a variant of rely-guarantee reasoning, adapted to
a replicated system where effectors execute atomically and definitively.
The precondition stability condition can be formally specified as follows:
\begin{small}
\begin{align}
\forall u_1, u_2 \in U,  \sigma, \sigma' \in \Sigma \centerdot \sigma \models (\Inv \wedge \Pre_{u_1} \wedge \Pre_{u_2}) \wedge u_1(\sigma) = \sigma' \implies \sigma' \models \Pre_{u_2}
\end{align}
\end{small}
\citet{syn:app:sh179} uses \emph{Tokens} to formalize concurrency control. 
Two operations that share the same token do not execute concurrently.
Since we are designing a coordination-free data structure, we consider the set of tokens to be an empty set, and hence absent from the formalisation.

\paragraph{Independence}
\label{sec:rules:concurrency:independency}

In order to ensure that the safety of an operation, is not impacted by skipping any previous operations, we augment the precondition stability analysis with an independence analysis as presented by \citet{Houshmand:2019:HRC:3302515.3290387}.
An operation $u_2$ is said to be independent of operation $u_1$ if the precondition of $u_2$, $\Pre_{u_2}$, is enabled even without executing $u_1$.
The condition for independency can be formally specified as follows:
\begin{small}
  \begin{align}
    \forall u_1, u_2 \in U,  \sigma, \sigma', \sigma'', \sigma''' \in \Sigma \centerdot \sigma \models (\Inv \wedge \Pre_{u_1}) \wedge \sigma' \models (\Inv \wedge \Pre_{u_2}) & \nonumber\\
    \wedge \sigma'' \models \Inv \wedge u_1(\sigma) = \sigma' \wedge u_2(\sigma') = \sigma'' \wedge u_2(\sigma) = \sigma''' & \implies \sigma \models \Pre_{u_2}  \wedge \sigma''' \models \Inv
  \end{align}
  \end{small}
In short, $u_2$ is independent of $u_1$ if, irrespective of whether $u_1$ executed before $u_2$, the execution of $u_2$ is safe.
This condition is required for safety only if the effect of $u_1$ is tentative, i.e., if $u_1$ has conflict resolution policies while applying the update on the state. 

\subsubsection{Mechanized verification}
\label{sec:rules:why3}
In order to mechanically discharge the proof obligations listed above,
we the use Why3 system \cite{filliatre:hal-00789533}, augmented with the
CISE3 plug-in \cite{meirim2020cise3}.
Why3 is a framework used for the deductive verification of programs,
\textit{i.e}., ``the process of turning the correctness of a program
into a mathematical statement and then proving it'' \cite{memoire}. 
The CISE3 plug-in automates the CISE proof rules described above, and
generates the required sequential-safety, commutativity and stability
checks.
Why3 then computes a set of proof obligations, that are discharged via
external theorem provers.

%% file: sections/incorrectspec.tex
\section{Sequential specification of a tree}
\label{sec:seqdesign}



\subsection{State} 
The state of a tree data structure consists of a set of nodes,
\emph{$\Nodes$}, and a relation from a child node to
its parent, indicated by $\rightarrow$.
The ancestor relation, $\rightarrow^*$ is defined as 
{\begin{small}

\begin{align}
  \forall a, n \in \Nodes \centerdot n \rightarrow^* a \implies & n \rightarrow a \ \vee \
  \exists p \in \Nodes \centerdot n \rightarrow p  \wedge p \rightarrow^* a \label{eq:ancestor}
\end{align}
\end{small}}
At initialization, the set of nodes consists of a single \emph{root} node.
The parent of the root is root itself.
The initial state of the tree is thus {\small$\Nodes = \{ \troot \}$} where {\small$\troot \rightarrow \troot$}. A crucial aspect of the abstract representation of the tree is how to express the relation between nodes.
Three choices are possible, either maintain a child-to-parent relation, a parent-to-child relation, or both. 
In particular, when implementing a tree, traversal efficiency depends on keeping both up and down pointers \cite{fic:rep:1787}.
Considering that child-to-parent and parent-to-child relations describe a dual view of a tree (i.e., node $p$ is 
the parent of node $n$ iff node $n$ is a descendent of node $p$) we selected the one that leads to a simpler 
specification. An advantage of using a child-to-parent relation is that it can be maintained as a function, as the
 tree properties ensure that each node has a unique parent. The alternative parent-to-child relation would 
 need a more complex representation, e.g. a function that maps each node to its set of direct descendants, 
 which would impact  the simplicity of the specification and the proof effort.
%


\subsection{Invariant} 
\label{sec:invariant}

The invariant of the tree data structure is as follows:
{\begin{small}

  \begin{align}
    \troot \in \Nodes \wedge \troot \rightarrow \troot \ \wedge\  
    \forall  \in \Nodes \centerdot n \neq \troot \implies \troot \not\rightarrow n  \label{eq:root}\tag{\textit{Root}} \nonumber\\[-2pt]
   \wedge\ \forall n \in \Nodes \centerdot \exists p \in \Nodes \centerdot n \rightarrow p  \label{eq:parent}\tag{\textit{Parent}}\nonumber\\[-2pt]
  \wedge\ \forall n,p,p' \in \Nodes \centerdot 
  n \rightarrow p \wedge n \rightarrow p'  \implies p = p' \label{eq:uniquep}\tag{\textit{Unique}}\nonumber\\[-2pt]
  \wedge\ \forall n \in \Nodes \centerdot  n \rightarrow^* \troot \label{eq:reachable}\tag{\textit{Reachable}}\nonumber\\[-2pt]
    \Inv \triangleq \ref{eq:root} \wedge \ref{eq:parent} \wedge \ref{eq:uniquep} \wedge \ref{eq:reachable} \label{eq:inv}
  \end{align}
\end{small}}
Clause \ref{eq:root} states that the root node is present in
$\Nodes$, and is the only node to be its own parent.
Clause \ref{eq:parent} asserts that every node in the tree has a parent in the tree.
Clause \ref{eq:uniquep} requires the parent of a node to be unique.
Clause \ref{eq:reachable} imposes that the root is an ancestor of all nodes.
We call this conjunction, \Cref{eq:inv}, the \emph{tree
  invariant}.

A further invariant
which forbids cycles 
can be derived:
{\begin{small}

  \begin{align*}
  \forall n \in \Nodes \centerdot n \neq \troot \implies n
  \not\rightarrow^* n \label{eq:acyclic}\tag{\textit{Acyclic}}
  \end{align*}
\end{small}}
\noindent{}
Since the parent relation inductively defines the ancestor relation, by \ref{eq:uniquep} there is a unique path to a given ancestor of a node.
By \ref{eq:reachable}, the root node is an ancestor of every node in the tree.
In this scenario, a cycle would require a node to have multiple parents, which is prevented by \ref{eq:uniquep}.



\subsection{Operations} 
\label{sec:operations}

\begin{description}[style=unboxed,leftmargin=0cm]
\item[{\color{black}Add}]
An add operation has two arguments: the node to be added, $n$, and its prospective parent, $p$.
The add effector adds node $n$ to $\Nodes$ and  the mapping $n \rightarrow p$ to the parent relation.
The postcondition of the add effector indicates this:%
\footnote{%
  For readability, we simplify the postcondition to express only the
  changes caused by the operation.
  The part of the state not mentioned remains unaffected.
}
{\begin{small}

\begin{align}
  \Post_{add(n,p)} \triangleq n \in \Nodes \wedge n \rightarrow p
\end{align}
\end{small}}
%
\end{description}
To ensure the tree invariant, we derive the
precondition that $n$ is a new node and $p$ is already in the tree, i.e.,
{\begin{small}

  \begin{align}
    \Pre_{add(n,p)} \triangleq n \notin \Nodes \wedge p \in \Nodes
  \end{align}
\end{small}}
%
%
\begin{table*}
  \begin{tabular}{@{\extracolsep{0pt}} c >{\centering\arraybackslash} p{1.45cm}  >{\centering\arraybackslash} p{2.85cm}  >{\centering\arraybackslash} p{1.5cm} >{\centering\arraybackslash} p{4.2cm}@{}}
    \toprule
    \multirow{2}{*}{{\normalsize Precondition}} & \multicolumn{4}{c}{{\normalsize Invariant clause}}\\
    \cline{2-5}
    & \emph{Root} & \emph{Parent} & \emph{Unique} & \emph{Reachable} \\
    \hline
    $add(n,p)$ & $n \notin \Nodes$ & $p \in \Nodes$ & $n \notin \Nodes$ & $p \in \Nodes $\\
    $rem(n)$ & $n \neq root$ & $\forall n' \in \Nodes \centerdot  n' \not\rightarrow n$ & $\mathtt{true}$ & $\forall n' \in \Nodes \centerdot n' \not\rightarrow n$ \\
    $move(n,p')$ & $n \neq root$ & $p' \in \Nodes$ & $\mathtt{true}$ & $p' \in \Nodes \wedge p' \neq n \wedge p' \not\rightarrow^* n$\\
    \bottomrule
  \end{tabular}
    \vspace{-5pt}
  \caption{Precondition required by each operation to uphold specific clauses of the invariant \vspace{-10pt}}
  \label{tab:wpcalculus}
\end{table*}
Let us see how this precondition is derived.
If the add operation is updating a safe state, i.e., the starting state
respects the invariant, and if the precondition is satisfied, then the
update should maintain the invariant.
Hereafter, we highlight the precondition clauses needed to ensure each
part of the invariant.%
\footnote{%
Denoted in inference style, as in \cite{syn:rep:lan:1814}.
The condition above the line represents the pre-state,
 an update event is noted $\llbracket.\rrbracket$, and the condition below the line indicates the post-state.
}
%
%
{\begin{small}

  \begin{align*}
    \inferrule
    {\Inv \wedge \focus{n \notin \Nodes} \\ \llbracket add(n, p) \rrbracket}
    {\Post_{add(n, p)} \wedge \ref{eq:root} \wedge \ref{eq:uniquep}}
    \qquad
    \inferrule
    {\Inv \wedge \focus{p \in \Nodes} \\  \llbracket add(n, p) \rrbracket}
    {\Post_{add(n, p)} \wedge \ref{eq:parent} \wedge \ref{eq:reachable} }
  \end{align*}
\end{small}}

%
\Cref{tab:wpcalculus} lists the preconditions required by operations to preserve each invariant clause.
With the derived preconditions, the add operation can be specified as follows:
%
{\begin{small}

  \begin{align*}
    \inferrule[(Add-Operation)]
    {\Inv \wedge n \notin \Nodes \wedge p \in \Nodes \\ \llbracket add(n, p)\rrbracket}
    {\Inv \wedge n \in \Nodes \wedge n \rightarrow p}
  \end{align*}
\end{small}}
  %
If the add operation is issued on a state that is safe and contains $p$ but not $n$, then $n$ is added to the tree with parent $p$.

\begin{description}[style=unboxed,leftmargin=0cm]
\item[{\color{black}Remove operation}]

Remove receives as argument a node $n$ to be deleted.
Its effector removes node $n$ from the set of nodes.
The postcondition of  remove  indicates this effect:
\end{description}
{\begin{small}

\begin{align}
  \Post_{rem(n)} \triangleq n \notin \Nodes
\end{align}
\end{small}}
Similarly to add, we list the predicates needed to preserve each clause of the invariant in \Cref{tab:wpcalculus}.
In the case of the remove operation, we need to ensure that $n$ is not the root, and $n$ is a leaf node, i.e., there are no child nodes for $n$.
%
%
%
The remove  can be specified as follows:
{\begin{small}

  \begin{align*}
    \inferrule[(Remove-Operation)]
    {\Inv \wedge n \neq \troot \wedge \forall \ n' \in \Nodes \centerdot n' \not\rightarrow n \\ \llbracket rem(n) \rrbracket}
    {\Inv \wedge n \notin \Nodes}
  \end{align*}
\end{small}}
If a remove operation is issued on a safe state where $n$ is not $\troot$ and has no children, then $n$ is removed from the tree.

\begin{description}[style=unboxed,leftmargin=0cm]
\item[{\color{black}Move operation}]
The move operation takes two arguments: the node to be moved $n$, and the new parent $p'$.
Its effector changes the parent of node $n$ to $p'$ as follows:
{\begin{small}
\begin{align}
  \Post_{move(n,p')} \triangleq n \rightarrow p'
\end{align}
\end{small}}
\end{description}
To preserve the expected behaviour we require that the node to be moved is already present in the tree. We derive the safety clauses as shown in \Cref{tab:wpcalculus}.
%
Formally, the move operation can be specified as follows:
{\begin{small}
  \begin{align*}
    \inferrule[(Move-Operation)]
    {\Inv \wedge n \in \Nodes \wedge n \neq \troot \wedge p' \in \Nodes \wedge p' \neq n \wedge p' \not\rightarrow^* n \\ \llbracket move(n,p') \rrbracket}
    {\Inv \wedge n \rightarrow p'}
  \end{align*}
\end{small}}
\noindent For the move operation to be safe, $n$ is not the root, $p'$ is in the tree, $n$ and $p'$ are different, and $p'$ is not a descendant of $n$.
These last two conditions are needed to prevent move from creating a
cycle of unreachable nodes, as we show with the following
counterexample.

Consider a tree composed of nodes \emph{a} and \emph{b}.
Root node \emph{R} is the parent of node \emph{a}, i.e., $a \rightarrow
R$ and node \emph{a} is the parent of node \emph{b}, $b \rightarrow a$,
and hence \emph{R} is the ancestor of \emph{b}, $b \rightarrow^* R$.
Moving  \emph{a} under  \emph{b} will make both \emph{a} and
\emph{b} unreachable from the root, and also form a cycle.
This violates the invariant by invalidating the tree structure.
To avoid this scenario, a precondition is needed that prevents moving a
node underneath itself.
When moving node $n$ from its current parent to the new parent
$p'$, $p'$ should not be a descendant of $n$, $p'
\not\rightarrow^* n$.

\subsection{Mechanized verification of the sequential specification}
\label{sec:cise3}
Following the formalization of the tree data structure above, we use
Why3 to mechanically prove its sequential safety.
The mechanical proof requires some extra definitions and axioms.

We need a predicate for reachability.
For this, we first define a path, a sequence of nodes related by the parent
relation.
We use $s[n]$ to indicate the nth element in the sequence $s$.
We denote the set of possible sequences of nodes by $S$.
The path predicate determines the validity conditions for a path $s$ between nodes $x$ and $y$ in state $\sigma$. 
If $x=y$, the path has length zero.
Otherwise, the length of the path is greater than zero, where the first path element must be $x$, all contiguous path elements are 
related by the parent relation, and node $y$ is the parent of  the last path element.
We say $y$ is reachable from $x$ if there exists a path from $x$ to $y$.
Formally, 
\begin{small}
  \begin{align}
    path(\sigma, x, y, s) \triangleq &\ length(s) = 0  \wedge x = y  \\
                             & \ \vee (length(s) > 0  \wedge s[0] = x  \wedge s[length(s)-1] \rightarrow y  \ \wedge \nonumber \\
                             & \hphantom{\wedge\wedge\wedge} \forall\, 0 \leq i < length(s)-1 \centerdot  s[i] \rightarrow s[i+1])  \nonumber\\
    reachability(\sigma, x, y) \triangleq &\exists s  \in S \centerdot path(\sigma, x, y, s)
  \end{align}
\end{small}
To formalize the properties of the \emph{path} predicate, we define a set of axioms as follows:
\begin{small}
  \begin{align}
    path\_to\_parent &\triangleq \forall \sigma \in \Sigma \centerdot \forall x,y \in \Nodes \centerdot x \rightarrow y  \implies \exists s \in S \centerdot path(\sigma, x,y,s) \wedge s = [x]\\
    path\_composition &\triangleq \forall \sigma \in \Sigma \centerdot \forall x,y,z \in \Nodes \centerdot \exists s_1\in S \centerdot path(\sigma, x,y,s_1)   \\ 
    & \hphantom{\wedge\wedge}  \wedge y \rightarrow z \implies \exists s_2 \in S \centerdot path(\sigma, x,z, s_2) \wedge s_2 = s_1 + [y] \nonumber \\
    path\_transitivity &\triangleq \forall \sigma \in \Sigma \centerdot \forall x,y,z \in \Nodes, s_1, s_2 \in S \centerdot path(\sigma, x,y,s_1)   \\ 
                     & \hphantom{\wedge\wedge} \wedge path(\sigma, y,z,s_2) \implies \exists s_3 \in S \centerdot path(\sigma, x,z,s_3) \wedge s_3 = s_1 + s_2 \nonumber\\
    path\_uniqueness &\triangleq \forall \sigma \in \Sigma \centerdot \forall x,y \in \Nodes, s_1,s_2 \in S \centerdot path(\sigma, x,y,s_1)   \\ 
                     & \hphantom{\wedge\wedge}  \wedge path(\sigma, x,y,s_2) \implies s_1 = s_2 \nonumber\\
    path\_exclusion &\triangleq \forall \sigma \in \Sigma \centerdot \forall x, y, z \in \Nodes, s \in S \centerdot x \not\rightarrow^* y  \wedge  path(\sigma, z,y,s) \implies x \notin s\\
    path\_separation &\triangleq \forall \sigma \in \Sigma \centerdot \forall x,y,z \in \Nodes, s_1,s_2 \in S   \centerdot  path(\sigma, x,y,s_1) \\ 
    & \hphantom{\wedge\wedge}  \wedge path(\sigma, y,z,s_2) \wedge  x \neq y \wedge x \neq z \wedge y \neq z \implies s_1 \cap s_2 = \emptyset \nonumber
  \end{align}
\end{small}
Axiom $path\_to\_parent$  defines the singleton path of a node to its
parent. The recursive composition of paths is axiomatized in $path\_composition$. 
The transitivity property is defined in $path\_transitivity$.
Axiom $path\_uniqueness$ asserts there is a single path between two nodes.
The  $path\_exclusion$ expresses the conditions for excluding nodes from a path.
Lastly, $path\_separation$ defines a convergence criterion essential for Why3's SMT solvers, 
asserting that the direction of the path is converging towards the root. 

We also require extra axioms to express the properties of the unaffected nodes in the case of add and move operations. 
They are as follows:
\begin{small}
  \begin{align}
    \sigma_{add} &= add(n, p)(\sigma)\nonumber\\
    \sigma_{move} &= move(n, p)(\sigma)\nonumber\\
    remaining\_nodes\_add &\triangleq \forall \sigma \in \Sigma \centerdot \forall n' \in \Nodes, s_1, s_2  \in \seqnodes \centerdot n' \neq n  
    \\ & \hphantom{\wedge\wedge}\wedge path(\sigma, n', \troot, s_1) \wedge path(\sigma_{add}, n', \troot, s_2) \implies  s_1 = s_2  \nonumber\\
    descendants\_move &\triangleq \forall \sigma \in \Sigma \centerdot  \forall n' \in \Nodes, s_1, s_2 \centerdot path(\sigma, n', c, s_1) 
    \\ & \hphantom{\wedge\wedge}\wedge path(\sigma_{move}, n', c, s_2) \implies s_1 = s_2 \nonumber\\
    remaining\_nodes\_move & \triangleq \sigma \in \Sigma \centerdot  \forall n' \in \Nodes, s_1, s_2 \centerdot n' \not\rightarrow^* n   
    \\ & \hphantom{\wedge \wedge } \wedge path(\sigma, n', \troot, s_1) \wedge path(\sigma_{move}, n', \troot, s_2) \implies s_1 = s_2 \nonumber
  \end{align}
\end{small}
The state $\sigma_{add}$ is obtained by applying $add(n,p)$ operation to $\sigma$.
The axiom $remaining\_nodes\_add$ asserts that the paths already present in the tree remain in the tree after executing the add operation.
Given that the move operation updates $\sigma$ to $\sigma_{move}$, axiom $descendants\_move$  asserts that the descendants of the node being moved continue to be its descendants, and $remaining\_nodes\_move$ asserts that other paths are not affected. These axioms are defined to ensure that the paths to the root, from nodes unaffected by move or add operations, remain unchanged. 
The specification proven using Why3  is available in \citet{tree_spec}.

%% file: sections/refinedspec.tex
\section{Concurrent tree specification}
\label{sec:refinedspec}

In this section, we discuss the convergence and concurrent safety of the tree.
In a sequential execution environment, as seen in \Cref{sec:seqdesign}, if the initial state and each individual update are safe, then all reachable states are safe.
This is not true when executing concurrently on multiple replicas.
In this case, there are three extra proof obligations (\Cref{sec:rules:concurrency:convergence}, \Cref{sec:rules:concurrency:stability}, \Cref{sec:rules:concurrency:independency}):

\begin{itemize}
\item   Ensuring that different replicas converge, despite effectors being executed concurrently in different orders.
\item   Ensuring that safety of an update is not violated by a concurrent update.
\item   Ensuring that a tentative update does not effect the safety of the dependent update.
\end{itemize}

For ease of exposition, 
first we discuss concurrent safety; convergence is deferred to
\Cref{sec:commutativity}, since the conflicts occurring in the latter can be addressed using the policies discussed in the former, and independence is discussed in \Cref{sec:independency}.

\subsection{Precondition stability}
\label{sec:tree:stability}

We use the precondition stability rule of CISE logic
(\Cref{sec:rules:concurrency:stability}) to analyze the concurrent safety of our tree data structure.
For each operation, we analyze whether it violates the precondition of
any other concurrent operation.
Formally, operation $op_1$ is stable under operation $op_2$ if, 
\begin{small}
	\begin{align}
		\inferrule
    {\Inv \wedge \Pre_{op_1} \wedge \Pre_{op_2} \\ \llbracket op_2 \rrbracket}
    {\Inv \wedge \Post_{op_2} \wedge \Pre_{op_1}}
		\label{eq:stability}
	\end{align}
\end{small}
We check the sequential specification for stability.
If this fails, then it will be necessary to modify the specification, so that it does satisfy stability.

\subsubsection{Stability of add operation}

\paragraph*{Concurrent adds:}
First we check the stability of the precondition of add against itself.
Let us consider two operations $add(n_1, p_1)$ and $add(n_2, p_2)$.
Using Equation~(\ref{eq:stability}), we get
\begin{small}
  \begin{align*}
    \Pre_{add(n_1, p_1)} &\triangleq n_1 \notin \Nodes \wedge p_1 \in \Nodes \nonumber\\
    \Pre_{add(n_2, p_2)} &\triangleq n_2 \notin \Nodes \wedge p_2 \in \Nodes \nonumber\\
    \Post_{add(n_2, p_2)} &\triangleq n_2 \in \Nodes \wedge n_2 \rightarrow p_2\nonumber
  \end{align*}
\end{small}
\vspace{-20pt}
\begin{small}
  \begin{align}
    \inferrule
    {\Inv \wedge \Pre_{add(n_1, p_1)} \wedge \Pre_{add(n_2, p_2)} \wedge \focus{n_1 \neq n_2} \\ \llbracket add(n_2, p_2) \rrbracket}
    {\Inv \wedge \Post_{add(n_2, p_2)} \wedge \Pre_{add(n_1, p_1)}}
  \end{align}
\end{small}
The highlighted clause $n_{1} \neq n_{2}$ is required for the stability
condition.
Indeed, the sequential specification does not disallow adding the same
node at different replicas, and the clause $n \notin \Nodes$ is unstable
therein.
Thus the analysis highlights a subtlety.

\paragraph*{Concurrent remove:}
Let us check the stability of the precondition of  $add(n_1, p_1)$ against a
concurrent $remove(n_2)$.
Using (\ref{eq:stability}), we get:
\begin{small}
  \begin{align*}
    \Pre_{add(n_1, p_1)} &\triangleq n_1 \notin \Nodes \wedge p_1 \in \Nodes \nonumber\\
    \Pre_{remove(n_2)} &\triangleq n_2 \neq \troot \wedge \forall \ n' \in \Nodes \centerdot n' \not\rightarrow n_2\nonumber\\
    \Post_{remove(n_2)} &\triangleq n_2 \notin \Nodes
  \end{align*}
\end{small}
\vspace{-20pt}
\begin{small}
  \begin{align}
    \inferrule
    {\Inv \wedge \Pre_{add(n_1, p_1)} \wedge \Pre_{remove(n_2)}
    \wedge \focus{n_2 \neq p_1} \\
    \llbracket remove(n_2) \rrbracket}
    {\Inv \wedge \Post_{remove(n_2)} \wedge \Pre_{add(n_1, p_1)}}
  \end{align}
\end{small}
In the sequential specification, clause $p_1 \in \Nodes$ in the
precondition of add is unstable against a remove of its parent;
performing those operations concurrently would be unsafe.

To fix this, we see two possible approaches.
The classical way is to strengthen the precondition with coordination,
for instance locking to avoid concurrency.
We reject this, as it conflicts with our objective of availability under
partition.
Our alternative is to weaken the specification thanks to
coordination-free conflict resolution.
We apply a common approach, to mark a node as deleted, as a so-called
\emph{tombstone}, without actually removing it from the data structure.%
\footnote{
  Ideally, one will remove the tombstone at some safe time in the
  future; this is non-trivial \cite{syn:1760} and out of the scope of
  this paper.
}

We now distinguish a \emph{concrete} state and its \emph{abstract} view.
We modify the specification to include a set of
tombstones, $\TS$ (initially empty), in the concrete state.
The abstract state is the resolved state as seen by some application
using \RDTName{}.
An \emph{abstraction function} maps the concrete state to the abstract state.

The concrete and abstract states of a tree are the same if either there are no nodes in the set of tombstones or for each node in the set of tombstones, all its descendants are also present in the set of tombstones.
In other cases, the abstraction function need to provide guidance on the presence of the descendants of a node that appears in the set of tombstones.


We present two \emph{abstraction functions} - \emph{skipping\_abstraction} and \emph{keeping\_abstraction}. 
The \emph{skipping\_abstraction} skips the descendants of the node that is marked as a tombstone.
The \emph{keeping\_abstraction}, on the other hand, preserves the tombstoned node if it observes the node has a descendant not in the set of tombstones.
Both the abstraction functions satisfy the required safety properties since they only change the view of the tree for an application.
Therefore the choice is application-specific.

Formally, if $\Nodes_{con}$ and $\Nodes_{abs}$ denote the set of nodes
in the concrete and abstract state respectively,
\begin{small}
	\begin{align}
		skipping\_abstraction \triangleq &
		\forall n \in \Nodes_{con} \centerdot n \notin \TS\ \wedge \not\exists n' \in \Nodes_{con} \centerdot \nonumber\\
		& \hphantom{\wedge\wedge\wedge\wedge} n' \in \TS\ \wedge n \rightarrow^* n' \Longleftrightarrow n \in \Nodes_{abs} \\
		keeping\_abstraction \triangleq &
		\forall n \in \Nodes_{con} \centerdot n \notin \TS\ \vee \exists n' \in \Nodes_{con} \centerdot \nonumber\\
		& \hphantom{\wedge\wedge\wedge\wedge}n' \notin \TS\ \wedge n' \rightarrow^* n \Longleftrightarrow n \in \Nodes_{abs}
  \end{align}
\end{small}
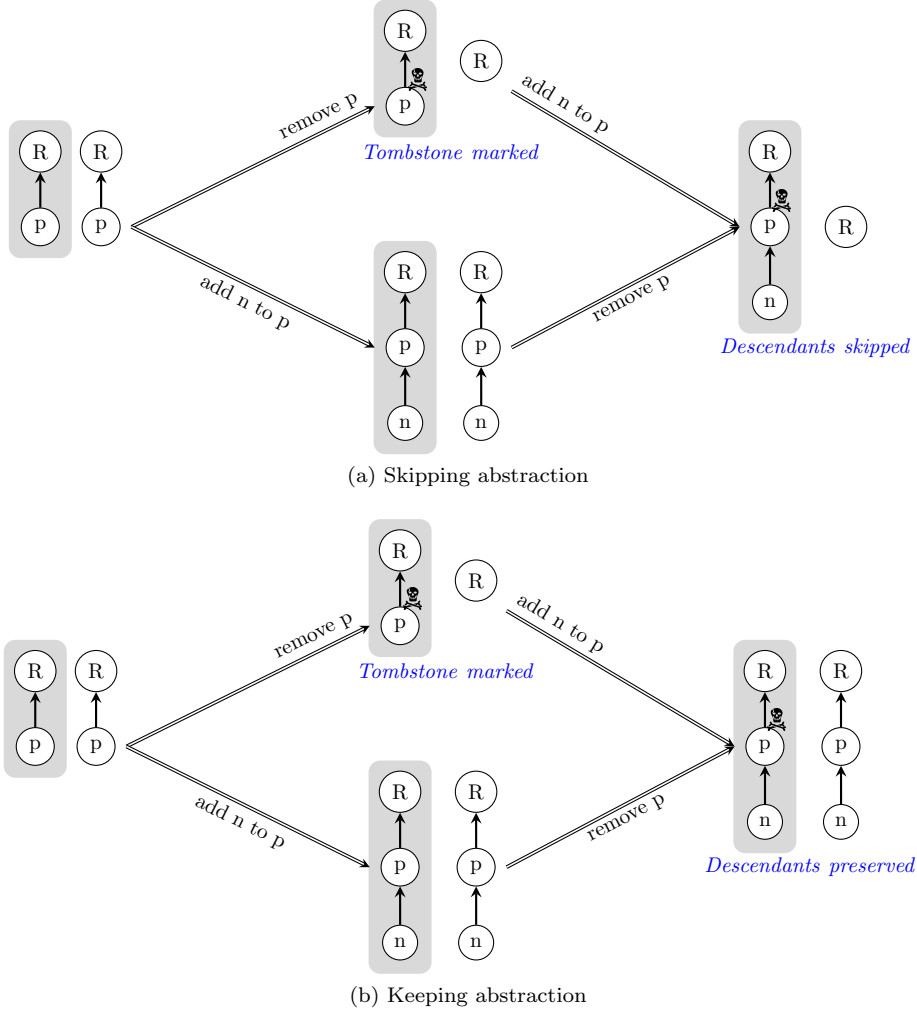
\begin{figure}[t]
  \subfloat[Skipping abstraction]{
  \centering
    \begin{tikzpicture}[scale=0.8, every node/.style={transform shape}]
      \tikzstyle{n} = [circle, minimum width=10pt, text centered, draw=black,fill=white]
      \tikzstyle{nodelist} = [circle, minimum width=18pt, text centered, draw=black,dashed,fill=white]
      
      \tikzstyle{arrow} = [thick,->,>=stealth]
      \tikzstyle{listarrow} = [dashed,->,>=stealth]
      \tikzstyle{operation} = [>=stealth,->,double]
      \tikzstyle{replica} = [>=open triangle 45,->,thin]
      
      \draw[rounded corners,gray!30,thick,fill=gray!30] (-0.5,0.5) rectangle (0.5,-1.75);
      \draw[rounded corners,gray!30,thick,fill=gray!30] (5.5,2.5) rectangle (6.5,0.25);
      \draw[rounded corners,gray!30,thick,fill=gray!30] (11.5,0.5) rectangle (12.5,-3);
      \draw[rounded corners,gray!30,thick,fill=gray!30] (5.5,-1.5) rectangle (6.5,-5);
      
      \node (root1) [n] {R};
      \node (p1) [n, below of=root1, yshift=-0.25cm] {p};
      
      \draw[arrow] (p1) -- (root1);
      
      \node (root1a) [n, right of=root1] {R};
      \node (p1a) [n, below of=root1a, yshift=-0.25cm] {p};
      
      \draw[arrow] (p1a) -- (root1a);
      
      \node (root2) [n, right of=root1, xshift=5cm, yshift=2cm] {R};
      \node [label={[xshift=0.2cm, yshift=-0.2cm]$\skull$}] (p2) [n, below of=root2, yshift=-0.25cm] {p};
      
      \draw[arrow] (p2) -- (root2);
      \node (root2a) [n, right of=root2, xshift=0.25cm, yshift=-0.5cm] {R};
      
      \node (root3) [n, right of=root1, xshift=5cm, yshift=-2cm] {R};
      \node (p3) [n, below of=root3, yshift=-0.25cm] {p};
      \node (n3) [n, below of=p3, yshift=-0.25cm] {n};
      
      \draw[arrow] (p3) -- (root3);
      \draw[arrow] (n3) -- (p3);
      
      \node (root3a) [n, right of=root3, xshift=0.25cm] {R};
      \node (p3a) [n, below of=root3a, yshift=-0.25cm] {p};
      \node (n3a) [n, below of=p3a, yshift=-0.25cm] {n};
      
      \draw[arrow] (p3a) -- (root3a);
      \draw[arrow] (n3a) -- (p3a);
      
      \node (root4) [n, right of=root2, xshift=5cm, yshift=-2cm] {R};
      \node [label={[xshift=0.2cm, yshift=-0.2cm]$\skull$}] (p4) [n, below of=root4, yshift=-0.25cm] {p};
      \node (n4) [n, below of=p4, yshift=-0.25cm] {n};
      
      \draw[arrow] (p4) -- (root4);
      \draw[arrow] (n4) -- (p4);
      
      \node (root4a) [n, right of=p4, xshift=0.25cm] {R};
      
      
      \coordinate[right of=p1a, xshift=-0.5cm] (from);
      \coordinate[left of=p2, xshift=0.5cm] (o1to);
      \draw [operation] (from) -- node[auto,sloped,pos=0.8] {remove p} (o1to);
      \coordinate[left of=p3, xshift=0.5cm] (o2to);
      \draw [operation] (from) -- node[auto,sloped,pos=0.5,below] {add n to p} (o2to);
      
      \coordinate[right of=root2a, xshift=-0.5cm, yshift=-0.5cm] (o1from);
      \coordinate[left of=p4, xshift=0.5cm] (to);
      \draw [operation] (o1from) -- node[auto,sloped,pos=0.2] {add n to p} (to);
      \coordinate[right of=p3a, xshift=-0.5cm] (o2from);
      \draw [operation] (o2from) -- node[auto,sloped,pos=0.5,below] {remove p} (to);
      
      \node [below of=root2, yshift=-1cm, xshift=0.75cm] {$\tabfocus{\mathit{Tombstone\ marked}}$};
      \node [below of=root4, yshift=-2.25cm, xshift=0.75cm] {$\tabfocus{\mathit{Descendants\ skipped}}$};
      
    \end{tikzpicture}
    \label{fig:remove-add:skipping}
  }\\
    \subfloat[Keeping abstraction]{
    \centering
    \begin{tikzpicture}[scale=0.8, every node/.style={transform shape}]
      \tikzstyle{n} = [circle, minimum width=10pt, text centered, draw=black,fill=white]
      \tikzstyle{nodelist} = [circle, minimum width=18pt, text centered, draw=black,dashed,fill=white]
      
      \tikzstyle{arrow} = [thick,->,>=stealth]
      \tikzstyle{listarrow} = [dashed,->,>=stealth]
      \tikzstyle{operation} = [>=stealth,->,double]
      \tikzstyle{replica} = [>=open triangle 45,->,thin]
      
      \draw[rounded corners,gray!30,thick,fill=gray!30] (-0.5,0.5) rectangle (0.5,-1.75);
      \draw[rounded corners,gray!30,thick,fill=gray!30] (5.5,2.5) rectangle (6.5,0.25);
      \draw[rounded corners,gray!30,thick,fill=gray!30] (11.5,0.5) rectangle (12.5,-3);
      \draw[rounded corners,gray!30,thick,fill=gray!30] (5.5,-1.5) rectangle (6.5,-5);
      
      \node (root1) [n] {R};
      \node (p1) [n, below of=root1, yshift=-0.25cm] {p};
      
      \draw[arrow] (p1) -- (root1);
      
      \node (root1a) [n, right of=root1] {R};
      \node (p1a) [n, below of=root1a, yshift=-0.25cm] {p};
      
      \draw[arrow] (p1a) -- (root1a);
      
      \node (root2) [n, right of=root1, xshift=5cm, yshift=2cm] {R};
      \node [label={[xshift=0.2cm, yshift=-0.2cm]$\skull$}] (p2) [n, below of=root2, yshift=-0.25cm] {p};
      
      \draw[arrow] (p2) -- (root2);
      \node (root2a) [n, right of=root2, xshift=0.25cm, yshift=-0.5cm] {R};
      
      \node (root3) [n, right of=root1, xshift=5cm, yshift=-2cm] {R};
      \node (p3) [n, below of=root3, yshift=-0.25cm] {p};
      \node (n3) [n, below of=p3, yshift=-0.25cm] {n};
      
      \draw[arrow] (p3) -- (root3);
      \draw[arrow] (n3) -- (p3);
      
      \node (root3a) [n, right of=root3, xshift=0.25cm] {R};
      \node (p3a) [n, below of=root3a, yshift=-0.25cm] {p};
      \node (n3a) [n, below of=p3a, yshift=-0.25cm] {n};
      
      \draw[arrow] (p3a) -- (root3a);
      \draw[arrow] (n3a) -- (p3a);
      
      \node (root4) [n, right of=root2, xshift=5cm, yshift=-2cm] {R};
      \node [label={[xshift=0.2cm, yshift=-0.2cm]$\skull$}] (p4) [n, below of=root4, yshift=-0.25cm] {p};
      \node (n4) [n, below of=p4, yshift=-0.25cm] {n};
      
      \draw[arrow] (p4) -- (root4);
      \draw[arrow] (n4) -- (p4);
      
      \node (root4a) [n, right of=root4, xshift=0.25cm] {R};
      \node (p4a) [n, below of=root4a, yshift=-0.25cm] {p};
      \node (n4a) [n, below of=p4a, yshift=-0.25cm] {n};
      
      \draw[arrow] (p4a) -- (root4a);
      \draw[arrow] (n4a) -- (p4a);
      
      \coordinate[right of=p1a, xshift=-0.5cm] (from);
      \coordinate[left of=p2, xshift=0.5cm] (o1to);
      \draw [operation] (from) -- node[auto,sloped,pos=0.8] {remove p} (o1to);
      \coordinate[left of=p3, xshift=0.5cm] (o2to);
      \draw [operation] (from) -- node[auto,sloped,pos=0.5,below] {add n to p} (o2to);
      
      \coordinate[right of=root2a, xshift=-0.5cm, yshift=-0.5cm] (o1from);
      \coordinate[left of=p4, xshift=0.5cm] (to);
      \draw [operation] (o1from) -- node[auto,sloped,pos=0.2] {add n to p} (to);
      \coordinate[right of=p3a, xshift=-0.5cm] (o2from);
      \draw [operation] (o2from) -- node[auto,sloped,pos=0.5,below] {remove p} (to);
      
      \node [below of=root2, yshift=-1cm, xshift=0.75cm] {$\tabfocus{\mathit{Tombstone\ marked}}$};
      \node [below of=root4, yshift=-2.25cm, xshift=0.75cm] {$\tabfocus{\mathit{Descendants\ preserved}}$};
      
    \end{tikzpicture}
    \label{fig:remove-add:keeping}
    }
    \caption{Resolving conflict of concurrent remove and add}
  \label{fig:remove-add}
\end{figure}
To illustrate the difference, consider the tree consisting of the root and a
single child, as shown in \Cref{fig:remove-add}.
One replica performs a remove of node $p$, while concurrently another
replica adds $n$ under $p$.
In the first replica, node $p$ is marked as a tombstone in the concrete
state (the shaded box).
Thus, the abstract state shows node $p$ removed.
When the replicas exchange their updates, they converge to the concrete state (the state in the shaded box).
\Cref{fig:remove-add:skipping} and \Cref{fig:remove-add:keeping} show the result of a \emph{skipping\_abstraction} and \emph{keeping\_abstraction} respectively.
In both the cases, node $p$ is marked as a tombstone.
In the case of the \emph{skipping\_abstraction}, node $p$ and the descendants are ``skipped''.
Meanwhile for \emph{keeping\_abstraction}, since
its descendant $n$ is not a tombstone, $p$ is ``revived'' in the
abstract view.

With tombstones, let us update the postcondition for remove:
\begin{small}
  \begin{align}
    \Post_{remove(n)} \triangleq n \in \TS
  \end{align}
\end{small}
Let us now derive the predicates needed to preserve each clause of the
invariant in this refined case.
\begin{small}
  \begin{align*}
    \inferrule
    {\Inv \wedge \focus{n \neq \troot} \\ \llbracket remove(n) \rrbracket}
    {\Post_{remove(n)} \wedge \ref{eq:root}}
    \qquad
    \inferrule
    {\Inv \wedge \focus{\mathtt{true}} \\ \llbracket remove(n) \rrbracket}
    {\Post_{remove(n)} \wedge \ref{eq:parent}}
    \\
    \inferrule
    {\Inv \wedge \focus{\mathtt{true}} \\ \llbracket remove(n) \rrbracket}
    {\Post_{remove(n)} \wedge \ref{eq:uniquep}}
    \qquad
    \inferrule
    {\Inv \wedge \focus{\mathtt{true}} \\ \llbracket remove(n) \rrbracket}
    {\Post_{remove(n)} \wedge \ref{eq:reachable}}
  \end{align*}
\end{small}
To maintain sequential safety in the modified remove specification, the
precondition forbids only removing the root node.
As the remove operation does not alter the tree structure, reachability
is not impacted.
The refined specification of the remove operation is as follows:
\begin{small}
  \begin{align*}
    \inferrule[(Remove-Operation)]
    {\Inv \wedge n \neq \troot \\ \llbracket remove(n) \rrbracket}
    {\Inv \wedge n \in \TS}
  \end{align*}
\end{small}

The application could strengthen this precondition with an added clause to delete only the leaf nodes visible in the abstract view. 
This helps prevent accident loss of a sub-tree.
Since this is not necessary for safety, we are not considering that condition.

\paragraph*{Concurrent move:}
Next we check the stability of the precondition of add under a concurrent move operation.
Let us consider two operations $add(n_1, p_1)$ and $move(n_2, p_2')$.
Using (\ref{eq:stability}), we get
\begin{small}
	\begin{align*}
		\Pre_{add(n_1, p_1)} &\triangleq n_1 \notin \Nodes \wedge p_1 \in \Nodes \nonumber\\
		\Pre_{move(n_2, p_2')} &\triangleq n_2 \in \Nodes \wedge n_2 \neq \troot \wedge p_2' \in \Nodes \wedge p_2' \neq n_2 \wedge p_2' \not\rightarrow^* n_2\nonumber\\
		\Post_{move(n_2, p_2')} &\triangleq n_2 \rightarrow p_2'
	\end{align*}
\end{small}\vspace{-12pt}
\begin{small}
	\begin{align}
		\inferrule
    {\Inv \wedge \Pre_{add(n_1, p_1)} \wedge \Pre_{move(n_2, p_2')} \wedge \focus{\mathtt{true}} \\ \llbracket move(n_2, p_2') \rrbracket}
    {\Inv \wedge \Post_{move(n_2, p_2')} \wedge \Pre_{add(n_1, p_1)}}
	\end{align}
\end{small}
We see that the precondition of add is stable against a concurrent move operation.

\subsubsection{Stability of remove operation}

\paragraph*{Concurrent add:}
Consider the sequential specification of two operations $remove(n_1)$ and $add(n_2, p_2)$.
Using (\ref{eq:stability}), we get
\begin{small}
	\begin{align*}
		\Pre_{remove(n_1)} &\triangleq n_1 \neq \troot \wedge \forall \ n' \in \Nodes \centerdot n' \not\rightarrow n_1\nonumber\\
		\Pre_{add(n_2, p_2)} &\triangleq n_2 \notin \Nodes \wedge p_2 \in \Nodes \nonumber\\
		\Post_{add(n_2, p_2)} &\triangleq n_2 \in \Nodes \wedge n_2 \rightarrow p_2
	\end{align*}
\vspace{-12pt}
	\begin{align}
		\inferrule
    {\Inv \wedge \Pre_{remove(n_1)} \wedge \Pre_{add(n_2, p_2)} \wedge \focus{n_1 \neq p_2} \\ \llbracket add(n_2, p_2) \rrbracket}
    {\Inv \wedge \Post_{add(n_2, p_2)} \wedge \Pre_{remove(n_1)}}
	\end{align}
\end{small}
We see that the clause that node $n_1$ has to be a leaf node is not satisfied if \mbox{$n_1 = p_2$} since add operation introduces a child node under $p_2$.
However, the refined specification of tombstones as described above does not require the node $n_1$ to be a leaf node.
So that solution fixes this conflict as well.

\paragraph*{Concurrent remove:}
Consider the sequential specification of two remove operations $remove(n_1)$ and $remove(n_2)$.
Using (\ref{eq:stability}), we get
\begin{small}
	\begin{align*}
		\Pre_{remove(n_1)} &\triangleq n_1 \neq \troot \wedge \forall \ n' \in \Nodes \centerdot n' \not\rightarrow n_1\nonumber\\
		\Pre_{remove(n_2)} &\triangleq n_2 \neq \troot \wedge \forall \ n' \in \Nodes \centerdot n' \not\rightarrow n_2\nonumber\\
		\Post_{remove(n_2)} &\triangleq n_2 \notin \Nodes
	\end{align*}
\vspace{-20pt}
\begin{align}
	\inferrule
    {\Inv \wedge \Pre_{remove(n_1)} \wedge \Pre_{remove(n_2)} \wedge \focus{\mathtt{true}} \\ \llbracket remove(n_2) \rrbracket}
    {\Inv \wedge \Post_{remove(n_2)} \wedge \Pre_{remove(n_1)}}
	\end{align}
\end{small}
We see that the remove operation is stable under a concurrent remove.
Furthermore, the refined specification is also stable since it adds $n_1$ and $n_2$ to $\TS$.

\paragraph*{Concurrent move:}
Consider the sequential specification of two operations $remove(n_1)$ and $move(n_2, p_2')$.
Using (\ref{eq:stability}), we get
\begin{small}
	\begin{align*}
		\Pre_{remove(n_1)} &\triangleq n_1 \neq \troot \wedge \forall \ n' \in \Nodes \centerdot n' \not\rightarrow n_1\nonumber\\
		\Pre_{move(n_2, p_2')} &\triangleq n_2 \in \Nodes \wedge n_2 \neq \troot \wedge p_2' \in \Nodes \wedge p_2' \neq n_2 \wedge p_2' \not\rightarrow^* n_2\nonumber\\
		\Post_{move(n_2, p_2')} &\triangleq n_2 \rightarrow p_2'
	\end{align*}
	\end{small}\vspace{-20pt}
	\begin{small}
	\begin{align}
		\inferrule
			{\Inv \wedge \Pre_{remove(n_1)} \wedge \Pre_{move(n_2, p_2')} \wedge \focus{n_1 \neq p_2'} \\ \llbracket move(n_2, p_2') \rrbracket}
			{\Inv \wedge \Post_{move(n_2, p_2')} \wedge \Pre_{remove(n_1)}}
	\end{align}
\end{small}
We see that the clause for the remove operation that $n_1$ should be a leaf node is violated if a node is moved under it.
Again, we can observe that the refined specification of remove eliminates this issue due to the absence of the violation-causing clause.

\subsubsection{Stability of move operation}

\paragraph*{Concurrent add:}
Consider the sequential specification of two operations $move(n_1, p_1')$ and $add(n_2, p_2)$.
Using (\ref{eq:stability}), we get
\begin{small}
	\begin{align*}
		\Pre_{move(n_1, p_1')} &\triangleq n_1 \in \Nodes \wedge n_1 \neq \troot \wedge p_1' \in \Nodes \wedge p_1' \neq n_1 \wedge p_1' \not\rightarrow^* n_1\nonumber\\
		\Pre_{add(n_2, p_2)} &\triangleq n_2 \notin \Nodes \wedge p_2 \in \Nodes\nonumber\\
		\Post_{add(n_2, p_2)} &\triangleq n_2 \in \Nodes \wedge n_2 \rightarrow p_2
	\end{align*}
\end{small}\vspace{-20pt}
\begin{small}
\begin{align}
	\inferrule
    {\Inv \wedge \Pre_{move(n_1, p_1')} \wedge \Pre_{add(n_2, p_2)} \wedge \focus{\mathtt{true}} \\ \llbracket add(n_2, p_2) \rrbracket}
    {\Inv \wedge \Post_{add(n_2,p_2)} \wedge \Pre_{move(n_1, p_1')}}
	\end{align}
\end{small}
The precondition of move is stable against a concurrent add operation.

\paragraph*{Concurrent remove:}
Consider the sequential specification of two remove operations $move(n_1, p_1')$ and $remove(n_2)$.
Using (\ref{eq:stability}), we get
\begin{small}
	\begin{align*}
		\Pre_{move(n_1, p_1')} &\triangleq n_1 \in \Nodes \wedge n_1 \neq \troot \wedge p_1' \in \Nodes \wedge p_1' \neq n_1 \wedge p_1' \not\rightarrow^* n_1\nonumber\\
		\Pre_{remove(n_2)} &\triangleq n_2 \neq \troot \wedge \forall \ n' \in \Nodes \centerdot n' \not\rightarrow n_2\nonumber\\
		\Post_{remove(n_2)} &\triangleq n_2 \notin \Nodes
	\end{align*}
\end{small}\vspace{-20pt}
\begin{small}
\begin{align}
	\inferrule
    {\Inv \wedge \Pre_{move(n_1, p_1')} \wedge \Pre_{remove(n_2)} \wedge \focus{n_2 \neq p_1'} \\ \llbracket remove(n_2) \rrbracket}
    {\Inv \wedge \Post_{remove(n_2)} \\\wedge \Pre_{move(n_1, p_1')}}
	\end{align}
\end{small}
Observe here that removing $n_2$ violates the clause $p_1' \in \Nodes$ if $n_2$ and $p_1'$ are the same.
However, in our refined specification, the postcondition of remove is $n_2 \in \TS$, keeping the clause $p_1' \in \Nodes$ stable.

\paragraph*{Concurrent move:}
Consider the sequential specification of two operations $move(n_1, p_1')$ and $move(n_2, p_2')$.
Using (\ref{eq:stability}), we get
\begin{small}
	\begin{align*}
		\Pre_{move(n_1, p_1')} &\triangleq n_1 \in \Nodes \wedge n_1 \neq \troot \wedge p_1' \in \Nodes \wedge p_1' \neq n_1 \wedge p_1' \not\rightarrow^* n_1\nonumber\\
		\Pre_{move(n_2, p_2')} &\triangleq n_2 \in \Nodes \wedge n_2 \neq \troot \wedge p_2' \in \Nodes \wedge p_2' \neq n_2 \wedge p_2' \not\rightarrow^* n_2\nonumber\\
		\Post_{move(n_2, p_2')} &\triangleq n_2 \rightarrow p_2'
	\end{align*}
\end{small}\vspace{-20pt}
\begin{small}
\begin{align}
	\inferrule
    {\Inv \wedge \Pre_{move(n_1, p_1')} \wedge \Pre_{move(n_2, p_2')} \wedge \focus{p_1' \not\rightarrow^* n_2} \\ \llbracket move(n_2, p_2') \rrbracket}
    {\Inv \wedge \Post_{move(n_2, p_2')} \wedge \Pre_{move(n_1, p_1')}}
	\end{align}
\end{small}
We see here that a concurrent move of $p_1$ or an ancestor of $p_1$ invalidates the precondition clause $p_1' \not\rightarrow^* n_1$ that prevents a cycle from forming.
This is a subtle condition missed in many previous works \cite{DBLP:journals/corr/abs-1805-04263,formel:rep:syn:sh197,fic:rep:sh172}; hence it highlights the value of a formal analysis.
We discuss this condition in more detail in \Cref{sec:move-move} and explain how we refine the specification for stability.

\renewcommand{\arraystretch}{1.4}
\begin{table}[t]
 \begin{centering}
  \begin{tabular}{@{\extracolsep{3pt}} c c c  >{\centering\arraybackslash} p{2.7cm}  >{\centering\arraybackslash} p{2.7cm}  >{\centering\arraybackslash} p{2.7cm} @{}}
  	\toprule
    	\multicolumn{2}{c}{\multirow{2}{*}{\textbf{Stability}}} &  \multicolumn{3}{c}{Stable against concurrent operation}\\
	\cline{3-5}
	 & & $\mathit{add}(n_2, p_2)$ & $\mathit{remove}(n_2)$ & $\mathit{move}(n_2, p_2')$\\
     	\hline
			 \multirow{4}{*}{\rotatebox[origin=l]{90}{\hspace{0.25cm}Operations}}
		 & $\mathit{add}(n_1, p_1)$ 	& $\tabfocus{n_1 \neq n_2}$
							& $\tabfocus{p_1 \neq n_2}$
							& $\tabfocus{true}$ \\
	\cline{3-5}
	 	 & $\mathit{remove}(n_1)$ 	& $\tabfocus{n_1 \neq p_2}$
							& $\tabfocus{true}$
							&  $\tabfocus{n_1 \neq p_2'}$ \\
	\cline{3-5}
	  & $\mathit{move}(n_1, p_1')$
	 						&  $\tabfocus{true}$
							& $\tabfocus{p_1' \neq n_2}$
							& $\tabfocus{p_1' \not\rightarrow^* n_2}$  \\
   	\bottomrule
  \end{tabular}
  \caption{Stability analysis of sequential specification}
  \label{tab:cise_analysis}
   \end{centering}
\end{table}
\renewcommand{\arraystretch}{1}
\Cref{tab:cise_analysis} shows the summary of the stability analysis on the sequential specification discussed in \Cref{sec:seqdesign}.
A condition indicates that the precondition of the operation in that row is stable under the operation in the column under the condition. 

\subsection{Safety of concurrent moves}
\label{sec:move-move}
We closely examine how a move operation on a remote replica might affect the precondition of a concurrent move in the local replica.
Consider an operation $move(n, p')$.
In a sequential execution, precondition clause $p' \not\rightarrow^* n$
forbids moving a node under itself (which would cause a cycle).
However a concurrent move of $p'$ under $n$ will not preserve the precondition of the operation, $p' \not\rightarrow^* n$, resulting in a cycle.

This issue generalizes to $p'$ or its ancestor concurrently moving under $n$ or a descendant of $n$.
For easy reference, we call this move as a \emph{cycle-causing-concurrent-move}.
Observe that the precondition prevents an ancestor of $n$ moving under itself in sequential execution.
Therefore, only the ancestors of $p'$ that are not ancestors of $n$ would lead to a cycle.
We call this set of ancestors \emph{critical ancestors}, and the set of $n$ and its descendants \emph{critical descendants} as defined in Table~\ref{tab:criticaldefinitionmove}.
\renewcommand{\arraystretch}{1.1}
\begin{table}
  \begin{tabular}{@{\extracolsep{0pt}} c >{\centering\arraybackslash} p{5cm} @{}}
    \toprule
    {\normalsize Property name} & {\normalsize Definition} \\
    \hline
    $\mathit{critical\_ancestors}$ &  $\{a \in \Nodes \centerdot p' \rightarrow^* a \wedge n \not\rightarrow^* a\}$\\
		$\mathit{critical\_descendants}$ & $\{d \in \Nodes \centerdot d \rightarrow^* n\}$\\
    \bottomrule
  \end{tabular}
  \caption{Critical ancestors and critical descendants of $move(n, p')$ \vspace{-10pt}}
  \label{tab:criticaldefinitionmove}
\end{table}
\renewcommand{\arraystretch}{1}

%
%
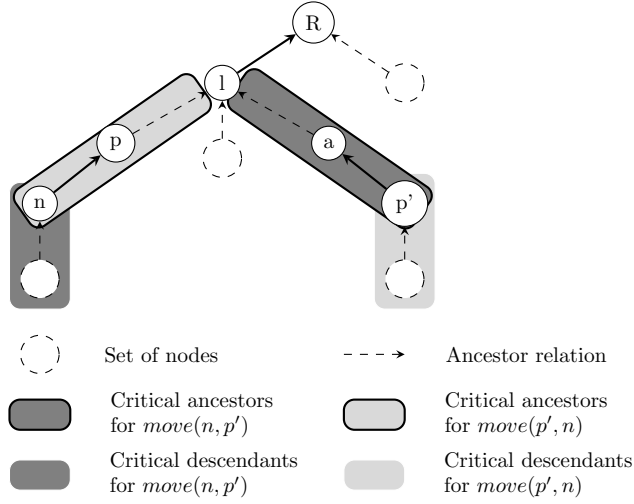
\begin{figure}[t]
  \centering
\begin{tikzpicture}[scale=0.8, every node/.style={transform shape}]
	\tikzstyle{n} = [circle, minimum width=10pt, text centered, draw=black,fill=white]
	\tikzstyle{nodelist} = [circle, minimum width=18pt, text centered, draw=black,dashed,fill=white]

	\tikzstyle{arrow} = [thick,->,>=stealth]
	\tikzstyle{listarrow} = [dashed,->,>=stealth]
	\tikzstyle{operation} = [>=stealth,->,double]

	\draw[rounded corners,white,fill=gray!100] (-5,-2.65) rectangle (-4,-4.75);
	\draw[rounded corners,white,fill=gray!30] (1,-2.5) rectangle (2,-4.75);

	\draw[rounded corners,black,thick,fill=gray!100,rotate around={55:(-1.5,0)}] (-2.6,-0.8) rectangle (-1.8,-4.5);
	\draw[rounded corners,black,thick,fill=gray!30,rotate around={125:(-1.5,0)}] (-2.58,0.9) rectangle (-1.83,4.5);

	\node (root) [n] {R};
	\node (lca) [n,below of=root, left of=root,xshift=-0.5cm] {l};
	\node (underroot) [nodelist,below of=root, right of=root,xshift=0.5cm] {};
	\node (underlca) [nodelist,below of=lca,yshift=-0.25cm] {};
	\node (p) [n,below of=lca, left of=lca,xshift=-0.75cm] {p};
	\node (a) [n,below of=lca, right of=lca,xshift=0.75cm] {a};
	\node (np) [n,below of=a, right of=a,xshift=0.25cm] {p'};
	\node (n) [n,below of=p,left of=p,xshift=-0.25cm] {n};
	\node (undern) [nodelist,below of=n,yshift=-0.25cm] {};
	\node (undernp) [nodelist,below of=np,yshift=-0.25cm] {};

	\draw [arrow] (lca) -- (root);
	\draw [listarrow] (underroot) -- (root);
	\draw [listarrow] (underlca) -- (lca);
	\draw [listarrow] (p) -- (lca);
	\draw [listarrow] (a) -- (lca);
	\draw [arrow] (np) -- (a);
	\draw [arrow] (n) -- (p);
	\draw [listarrow] (undern) -- (n);
	\draw [listarrow] (undernp) -- (np);

	\node (legend) [nodelist, below of=undern, yshift=-0.25cm] {};
	\node [align=left, right of=legend, xshift=1cm]  {Set of nodes};
	\draw [listarrow] (.5,-5.5) -- (1.5,-5.5);
	\node [align=left] at (3.5, -5.5) {Ancestor relation};
	\draw[rounded corners,black,thick,fill=gray!100] (-4,-6.25) rectangle (-5,-6.75);
	\node [align=left] at (-2, -6.5) {Critical ancestors \\for $move(n, p')$};
	\draw[rounded corners,white,fill=gray!100] (-4,-7.25) rectangle (-5,-7.75);
	\node [align=left] at (-1.8, -7.5) {Critical descendants \\for $move(n, p')$};
	\draw[rounded corners,black,thick,fill=gray!30] (0.5,-6.25) rectangle (1.5,-6.75);
	\node [align=left] at (3.5, -6.5) {Critical ancestors \\for $move(p', n)$};
	\draw[rounded corners,white,fill=gray!30] (0.5,-7.25) rectangle (1.5,-7.75);
	\node [align=left] at (3.7, -7.5) {Critical descendants \\for $move(p', n)$};
\end{tikzpicture}
\caption{Critical ancestors and critical descendants }
\label{fig:criticalcomp}
\end{figure}
Consider two concurrent move operations $move(n, p')$ and $move(p', n)$.
Figure~\ref{fig:criticalcomp} shows the critical ancestors and critical
descendants of both move operations. The node~$l$ is common ancestor of both $n$
and $p'$ farthest from the root. The critical ancestors and critical descendants
of $move(n, p')$ are grouped together in the dark gray region with and
without a border respectively, and that of $move(p', n)$ are grouped together
in the light gray region.

Note that the set of critical descendants of a move overlaps with the
critical ancestors set of its corresponding cycle-causing-concurrent-move.
Hence, we consider only the critical ancestors of move operations.

Let us take a step back and analyze the types of move operations.  Some move
operations result in a node moving farther away from the root, called
\emph{down-move}s, and another set of move operations result in the node moving
nearer to the root, or to remain at the same distance from the root, called
\emph{up-move}s.  We define $\rank$ as the distance of a
node from the root node, as follows:
\begin{small}
\begin{align}
  \rank(\troot) & = 0\\
	\rank(n) & = \rank(p) + 1 \mid \forall  n,p \in \Nodes \centerdot n \rightarrow p\\
		up\mhyphen move(n, p') & \implies rank(n) > rank(p')\\
		down\mhyphen move(n, p') & \implies rank(n) \leq rank(p')
	\end{align}
\end{small}
Consider a move operation, $move(n, p')$, moving node $n$ at the same level
or towards the root, i.e., an up-move.  This gives us that
$\rank(n) > \rank(p')$.  In this case, the rank of a critical descendant will be
always greater than the rank of a critical ancestor.  Formally,
\begin{small}
\begin{align}
	\forall n,p,p',d,a \in \Nodes \centerdot n \rightarrow p \wedge \rank(n) > \rank(p') \nonumber\\\wedge d \rightarrow^* n \wedge p' \rightarrow^* a 
	\implies \rank(d) > \rank(a)
\end{align}
\end{small}
This implies that a cycle-causing-concurrent-move can only be a down-move.  Hence, we
have that concurrent up-moves are safe; stability issues can occur only between
two concurrent down-moves, or between an up-move and a down-move.

Our next step is to design a coordination-free conflict resolution
policy for the moves that conflict.
The conflict resolution policy is required if both the concurrent
move operations move a node in the set of critical ancestors of the other.  If
we have up-moves, we apply the effect of the operation.  In case of a concurrent
down-move and up-move, up-move wins and the down-move is skipped.  In case of
concurrent down-moves, we apply a deterministic conflict resolution policy;
the operation with highest \emph{priority number} wins.  The priority number of
a move operation is specific to each application, with a condition that it must
be unique for each move.


Contrast our approach with the alternative that uses shared-exclusive
locks for concurrent moves \cite{formel:rep:syn:sh197}.
Consider concurrent operations $move(n, p')$, moving node $n$ under
$p'$, and $move(p', n)$, moving node $p'$ under $n$.
These operations compete for a lock.
The one that succeeds first will apply its move, blocking the other.
When it releases the lock, this releases the second one, but its
precondition is no longer valid and it cannot execute.
Thereby, safety is preserved, at the cost of aborting the second move.
This work essentially achieves the same end result, but without the
overhead of locking.
Our experiments in Section~\ref{sec:evaluation} show the performance
difference.

\subsection{Convergence}
\label{sec:commutativity}

\renewcommand{\arraystretch}{1.1}
\begin{table}[t]
  \begin{centering}
    \begin{tabular}{@{\extracolsep{0pt}}c >{\centering\arraybackslash} p{1.2cm}  >{\centering\arraybackslash} p{1.2cm}  >{\centering\arraybackslash} p{2.8cm} @{}}
      \toprule
      \multirow{2}{*}{{\normalsize \textbf{Commutativity}}} &  \multicolumn{3}{c}{{\normalsize Operations}}\\
      \cline{2-4}
        & $\!\!\!\mathit{add}(n_2, p_2)$ & $\ \mathit{rem}(n_2)$ & $\mathit{move}(n_2, p_2')$\\
      \hline
        $\mathit{add}(n_1, p_1)$ 	&  \textcolor{darkgreen}{\ding{52}}
            &  \textcolor{darkgreen}{\ding{52}}
            & \textcolor{darkgreen}{\ding{52}} \\[1pt]
        $\mathit{rem}(n_1)$ 	& \textcolor{darkgreen}{\ding{52}}
            & \textcolor{darkgreen}{\ding{52}}
            & \textcolor{darkgreen}{\ding{52}}  \\[1pt]
        $\mathit{move}(n_1, p_1')$
          & \textcolor{darkgreen}{\ding{52}}
            &  \textcolor{darkgreen}{\ding{52}}
            & \mbox{$\!\!\!\!\tabfocus{\lnot(n_1 = n_2 \wedge p_1' \neq p_2')}$} \\
      \bottomrule
    \end{tabular}
    \caption{Result of commutativity analysis of the sequential specification discussed in \Cref{sec:seqdesign} }
    \label{tab:commutativity}
  \end{centering}
\end{table}
\renewcommand{\arraystretch}{1}

As discussed in Section~\ref{sec:rules}, to ensure convergence, we design the
data structure such that concurrent updates commute~\cite{syn:rep:sh143}.
Add and remove operations result in adding the added and removed node to
$\Nodes$ and $\TS$ respectively.  Since set union is commutative, each
of these two operations commutes with itself and with the other.

The move operation changes the parent pointer of a node.  It commutes with add
and remove, since it doesn't have an effect on set membership.

However, observe that in the sequential specification two moves do not
commute, if the same node is moved to two different places.
This issue is fixed by the conflict resolution policy discussed earlier.
The results of the commutativity analysis is show in
Table~\ref{tab:commutativity}.

\subsection{Independence}
\label{sec:independency}

We use the independence conditions from \Cref{sec:rules:concurrency:independency} to check for safety violations due to tentative moves.
We check whether each operation is independent of up-move and down-move since they are the only operations that have tentative effects.
For the dependent operations, we compute the condition under which it is dependent and use it to devise dependency resolution policies.
In order to compute dependency conditions, we use the dependency analysis proposed by \citet{Houshmand:2019:HRC:3302515.3290387}.
An operation $op_2$ is dependent on $op_1$ if the execution of $op_1$ enabled $\Pre_{op_2}$ that was not enabled before its execution, i.e., 
\begin{small}
  \begin{align}
    \inferrule
    {\Inv \wedge \Pre_{op_1} \\ \llbracket op_1 \rrbracket}
    {\Inv \wedge \Post_{op_1} \wedge \Pre_{op_2}}
  \end{align}
  \label{eq:tree:independence}
\end{small}
If the dependency condition evaluates to $\mathtt{true}$, then $op_2$ is independent of $op_1$.
We use this analysis to check the independence of add, remove, up-move and down-move with respect to a historical tentative operation, i.e., an up-move or down-move performed before the operation under observation.

\subsubsection{Independence of add operation}
\paragraph*{Historical up-move:}
We use Equation~\ref{eq:tree:independence} as follows:
\begin{small}
  \begin{align*}
    \Pre_{up\mhyphen move(n_1,p_1')} &\triangleq n_1 \in \Nodes \wedge n_1 \neq \troot \wedge p_1' \in \Nodes  \wedge\,  n_1 \neq p_1' \wedge p_1' \not\rightarrow^* n_1 \wedge \rank(n_1) > \rank(p_1')\\
    \Post_{up\mhyphen move(n_1,p_1')} &\triangleq \mathtt{skip} \vee n_1 \rightarrow p_1'\\ 
    \Pre_{add(n_2,p_2)} &\triangleq p_2 \in \Nodes \wedge n_2 \notin \Nodes
  \end{align*}
\end{small}
Since the historical up-move doesn't change the membership of $\Nodes$, we can see that add is independent of up-move.

\paragraph*{Historical down-move:}
An add operation is independent of a historical down-move in the same manner because it does not change the membership of $\Nodes$ either.

\subsubsection{Independence of remove operation}
\paragraph*{Historical up-move:}
For checking the independence of remove, Equation~\ref{eq:tree:independence} becomes:
\begin{small}
  \begin{align*}
    \Pre_{up\mhyphen move(n_1,p_1')} &\triangleq n_1 \in \Nodes \wedge n_1 \neq \troot \wedge p_1' \in \Nodes  \wedge\,  n_1 \neq p_1' \wedge p_1' \not\rightarrow^* n_1 \wedge \rank(n_1) > \rank(p_1')\\
    \Post_{up\mhyphen move(n_1,p_1')} &\triangleq \mathtt{skip} \vee n_1 \rightarrow p_1'\\ 
    \Pre_{remove(n_2)} &\triangleq n_2 \neq \troot
  \end{align*}
\end{small}
Since $n_2 \neq \troot$ is unaffected by a historical up-move, remove is independent of up-move.

\paragraph*{Historical down-move:}
Similarly to historical up-move, a historical down-move also has no impact of the precondition of a remove operation.
Hence remove is independent of a historical down-move.

\subsubsection{Independence of up-move operation}
\paragraph*{Historical up-move:}
Now we analyse whether an up-move is independent of a historical up-move.
\begin{small}
  \begin{align*}
    \Pre_{up\mhyphen move(n_1,p_1')} &\triangleq n_1 \in \Nodes \wedge n_1 \neq \troot \wedge p_1' \in \Nodes  \wedge\,  n_1 \neq p_1' \wedge p_1' \not\rightarrow^* n_1 \wedge \rank(n_1) > \rank(p_1')\\
    \Post_{up\mhyphen move(n_1,p_1')} &\triangleq \mathtt{skip} \vee n_1 \rightarrow p_1'\\ 
    \Pre_{up\mhyphen move(n_2,p_2')} &\triangleq n_2 \in \Nodes \wedge n_2 \neq \troot \wedge p_2' \in \Nodes  \wedge\,  n_2 \neq p_2' \wedge p_2' \not\rightarrow^* n_2 \wedge \rank(n_2) > \rank(p_2')
  \end{align*}
\end{small}

We first divide the postcondition of the historical up-move into two parts: on the one hand, $\mathtt{skip}$, which leaves the state as it was; and on the other hand, $n_1 \rightarrow p_1'$, which changes the parent relation.
Then we divide the precondition of the second up-move into two parts, $n_2 \in \Nodes \wedge n_2 \neq \troot \wedge p_2' \in \Nodes  \wedge\,  n_2 \neq p_2'$, which is unaffected by the historical up-move, and $p_2' \not\rightarrow^* n_2 \wedge \rank(n_2) > \rank(p_2')$, which is potentially effected by the second part of the postcondition of the historical up-move.

Note that $\Pre_{up\mhyphen move(n_2,p_2')}$ was not enabled before the execution of $op_1$, i.e., the execution of $op_1$ enabled at least one predicate $p_2' \not\rightarrow^* n_2$ or $\rank(n_2) > \rank(p_2')$.
%
Let us consider them one at a time.

Let us derive the conditions under which moving a node to a different parent introduces an ancestor relation that enables the condition $p_2' \not\rightarrow^* n_2$ (it was previously disabled).
This means that the historical up-move operation caused a disconnection between $p_2'$ and $n_2$. 
This will happen only if the node being moved by the historical up-move was either $n_2$ or a descendant of $n_2$ and the new parent of the current move was either $n_1$ or a descendant of $n_1$ (the node moved by the historical up-move).
Hence we have $(n_1 = n_2 \vee n_1 \rightarrow^* n_2) \wedge (p_2' = n_1 \vee p_2' \rightarrow^* n_1)$.

The condition $\rank(n_2) > \rank(p_2')$ will be enabled after an up-move only if $\rank(p_2')$ decreased.\footnote{Note that $\rank(n_2)$ cannot increase since an up-move does not cause the rank of any move to increase.}
This will happen only if $p_2'$ was the node moved or its descendant.
Hence we have that $p_2' = n_1 \vee p_2' \rightarrow^* n_1$.

The historical up-move either enabled one or both of the conditions. 
Combining them gives $p_2' = n_1 \vee p_2' \rightarrow^* n_1$, the condition under which an up-move, $up\mhyphen move(n_2,p_2')$, is dependent on a historial up-move, $up\mhyphen move(n_1,p_1')$.

\paragraph*{Historical down-move:}
To check for independence of an up-move with a historical down-move, we have the following condition:
\begin{small}
  \begin{align*}
    \Pre_{down\mhyphen move(n_1,p_1')} &\triangleq n_1 \in \Nodes \wedge n_1 \neq \troot \wedge p_1' \in \Nodes  \wedge\,  n_1 \neq p_1' \wedge p_1' \not\rightarrow^* n_1 \wedge \rank(n_1) \leq \rank(p_1')\\
    \Post_{down\mhyphen move(n_1,p_1')} &\triangleq \mathtt{skip} \vee n_1 \rightarrow p_1'\\ 
    \Pre_{up\mhyphen move(n_2,p_2')} &\triangleq n_2 \in \Nodes \wedge n_2 \neq \troot \wedge p_2' \in \Nodes  \wedge\,  n_2 \neq p_2' \wedge p_2' \not\rightarrow^* n_2 \wedge \rank(n_2) > \rank(p_2')
  \end{align*}
\end{small}
We apply the same reasoning as in the previous case for the condition $p_2' \not\rightarrow^* n_2$, obtaining $(n_1 = n_2 \vee n_1 \rightarrow^* n_2) \wedge (p_2' = n_1 \vee p_2' \rightarrow^* n_1)$ as the condition under which an up-move is dependent under a historical down-move.

There is a difference in the second part though; the condition $\rank(n_2) > \rank(p_2')$ will be enabled after a down-move only if $\rank(n_2)$ increases (not possible for a down-move to decrease the rank).
This will happen only if $n_2$ was the node moved or its descendant, i.e., $n_2 = n_1 \vee n_2 \rightarrow^* n_1$.

Combining both the conditions, we have $((n_1 = n_2 \vee n_1 \rightarrow^* n_2) \wedge (p_2' = n_1 \vee p_2' \rightarrow^* n_1)) \vee (n_2 = n_1 \vee n_2 \rightarrow^* n_1)$ as the condition under which an up-move, $up-move(n_2,p_2')$, is dependent on a historial down-move, $down\mhyphen move(n_1,p_1')$.

\subsubsection{Independence of down-move operation}
\paragraph*{Historical up-move:}
The pre and postconditions required to analyse the dependence of a down-move operation under a historical up-move is as follows:
\begin{small}
  \begin{align*}
    \Pre_{up\mhyphen move(n_1,p_1')} &\triangleq n_1 \in \Nodes \wedge n_1 \neq \troot \wedge p_1' \in \Nodes  \wedge\,  n_1 \neq p_1' \wedge p_1' \not\rightarrow^* n_1 \wedge \rank(n_1) > \rank(p_1')\\
    \Post_{up\mhyphen move(n_1,p_1')} &\triangleq \mathtt{skip} \vee n_1 \rightarrow p_1'\\ 
    \Pre_{down\mhyphen move(n_2,p_2')} &\triangleq n_2 \in \Nodes \wedge n_2 \neq \troot \wedge p_2' \in \Nodes  \wedge\,  n_2 \neq p_2' \wedge p_2' \not\rightarrow^* n_2 \wedge \rank(n_2) \leq \rank(p_2')
  \end{align*}
\end{small}
Note that the reasoning for the up-move operation also remains valid here since the effect of both moves are the same, only their preconditions differ, only the clause comparing the ranks of the node and the new parent differs.
The first part of the dependency condition remains, $(n_1 = n_2 \vee n_1 \rightarrow^* n_2) \wedge (p_2' = n_1 \vee p_2' \rightarrow^* n_1)$.

The condition $\rank(n_2) \leq \rank(p_2')$ will be effected only if the historical up-move decreased the rank of $n_2$.
Hence we have the condition $n_2 = n_1 \vee n_2 \rightarrow^* n_1$.

Combining the clauses, we have $((n_1 = n_2 \vee n_1 \rightarrow^* n_2) \wedge (p_2' = n_1 \vee p_2' \rightarrow^* n_1)) \vee (n_2 = n_1 \vee n_2 \rightarrow^* n_1)$, the condition under which a down-move is dependent on a historical up-move.

\paragraph*{Historical down-move:}
We consider the following pre and postconditions:
\begin{small}
  \begin{align*}
    \Pre_{down\mhyphen move(n_1,p_1')} &\triangleq n_1 \in \Nodes \wedge n_1 \neq \troot \wedge p_1' \in \Nodes  \wedge\,  n_1 \neq p_1' \wedge p_1' \not\rightarrow^* n_1 \wedge \rank(n_1) \leq \rank(p_1')\\
    \Post_{down\mhyphen move(n_1,p_1')} &\triangleq \mathtt{skip} \vee n_1 \rightarrow p_1'\\ 
    \Pre_{down\mhyphen move(n_2,p_2')} &\triangleq n_2 \in \Nodes \wedge n_2 \neq \troot \wedge p_2' \in \Nodes  \wedge\,  n_2 \neq p_2' \wedge p_2' \not\rightarrow^* n_2 \wedge \rank(n_2) \leq \rank(p_2')
  \end{align*}
\end{small}
We use the reasoning as in the previous cases on these ang get $p_2' = n_1 \vee p_2' \rightarrow^* n_1$, the condition under which a down-move, $down\mhyphen move(n_2,p_2')$, is dependent on a historial down-move, $down\mhyphen move(n_1,p_1')$.

We see that up-move and down-move operations are dependent on each other and add and remove are independent of up-move and down-move.
We also derived the conditions under which up-moves and down-moves are dependent on each other.
We use this information to design dependence resolution policies.

\subsection{Safe specification of a replicated tree}
\label{sec:spec}
\begin{figure*}
  \begin{small}
    \begin{flushleft}
      \input{sections/concspec.tex}
    \end{flushleft}
  \end{small}
  \caption{Concurrent specification of \RDTName}
  \label{fig:concspec}
\end{figure*}

We incorporate the stability, commutativity, and independence analysis results and the
design refinements, resulting in the coordination-free, safe and convergent replicated tree
data structure specified in Specification~\ref{fig:concspec}.  
The state now consists of a set of nodes, $\Nodes$, and tombstones, $\TS$.  Since the tombstones also form
part of the tree, they also have to maintain the tree structure.  
The invariants refer to the set of nodes which includes tombstones.

We also introduce some definitions to help define the coordination-free and
conflict-free up-move and down-move operations.  We define an operation as a
tuple consisting of its type (add, remove, up-move or down-move), its parameters, and its
priority. 
The priority is arbitrary (e.g. supplied by the application);
the only condition being that priorities are totally ordered.
We define $\mathtt{C}$ as the set of operations concurrent with the
operation under consideration.  
$\mathtt{H}$ is the set of operations seen by the current operation.
We also define operations on critical ancestors as
$crit\mhyphen anc\mhyphen overlap$, where the node being moved is a member of the set of
critical ancestors of the other operation.
$\mathit{self\mhyphen or\mhyphen under}$ indicates the node itself and its descendants.

With the help of these definitions, we define the up-move and
down-move operations in three parts: the actual precondition needed to
ensure sequential safety, the conflict resolution condition
(highlighted in light blue), the dependency condition (highlighted in dark blue), and the update on the state.
Note that the conflict resolution and dependency checks are performed while applying the effect of the operation on the local and remote replicas, while the precondition is checked only at the local replica.

\subsection{Mechanized verification of the concurrent specification}
We use the CISE3 plug-in, presented in Section~\ref{sec:rules:why3}, to
identify conflicts as shown in Tables~\ref{tab:cise_analysis}
and~\ref{tab:commutativity}.
Given the sequential
specification from Section~\ref{sec:seqdesign}, CISE3 automatically generates a set
of meta-operations to check stability and commutativity of executing pairs of
operations. 

\subsubsection{Provable concurrent execution} 
We update the Why3 specification according to
the conflict resolution policies from Section \ref{sec:spec}. 
For example, for the add operation
we place the new
precondition that nodes must be uniquely identified:
\begin{small}
\begin{why3}
  assume { ... /\ n1 <> n2 }
\end{why3}
\end{small}
%
Next, we refine the definition of type {\small\texttt{state}} to include
tombstones, as follows:
\begin{small}
\begin{why3}
  type state = { mutable nodes: fset elt; ...;
  	mutable tombstones: fset elt; }
  \end{why3}
\end{small}
%
We update the specification of the {\small\texttt{rem}} accordingly:
\begin{small}
\begin{why3}
  val rem (n : elt) (s : state) : unit
    ensures  { s.tombstones = add n (old s).tombstones }
\end{why3}
\end{small}
where {\small\texttt{add}} stands for the logical adding operation on sets.

Finally, the implementation of the conflict resolution policy for a pair of
{\small\texttt{move}} operations requires us to be a bit more creative. 
We update the
{\small\texttt{state}} type definition to include ranking and critical ancestors
information.
We implement a custom analysis {\small\texttt{move\_refined}}
operation 
since concurrent operations are not available off-the-shelf in Why3, a
framework for verification of sequential specifications.
We encode the
arguments of two move operations as arguments of the {\small\texttt{move\_refined}}
operation: {\small\texttt{n1}}
({\small\texttt{n2}}), {\small\texttt{np1}} ({\small\texttt{np2}}), and {\small\texttt{pr1}} ({\small\texttt{pr2}}) stand for the
node to be moved, the new parent, and the unique priority levels respectively, of the first (second)
move. 

All analysis functions, except \texttt{move\_refined}, 
 are automatically generated by the CISE3
plug-in of Why3. 
Finally, 55 verification
conditions are generated for the implementation and given specification of
\texttt{move\_refined}. All of these are automatically verified, using a
combination of SMT solvers.
The specification and the proof results are available at \cite{tree_spec}.

%% file: sections/concspec.tex



\textbf{State:}
\begin{small}
$\Nodes \times \TS$
\end{small}

\textbf{Invariant:}
\vspace{-20pt}
\begin{small}
  \begin{align*}
  \troot \rightarrow \troot \wedge \forall n \in \Nodes \centerdot &\ \troot \not\rightarrow  n \wedge \troot \notin \TS\label{eq:croot}\tag{\textit{Root}}\\
  \wedge\ \forall n \in \Nodes \centerdot &\ n \neq \troot \wedge \exists p \in \Nodes \centerdot n \rightarrow p \label{eq:cparent}\tag{\textit{Parent}}\\
  \wedge\ \forall n,p,p' \in \Nodes \centerdot 
  n \rightarrow p \wedge n \rightarrow p' & \implies p = p' \label{eq:cuniquep}\tag{\textit{Unique}}\\
  \wedge\ \forall n \in \Nodes \centerdot n \neq \troot & \implies n \rightarrow^* \troot \label{eq:creachable}\tag{\textit{Reachable}}\\[-20pt]
\end{align*}  
\end{small}
\begin{minipage}{0.45\textwidth}
  \textbf{Add operation:}
  \begin{small}
    \begin{align*}
      \inferrule[(Add-Operation)]
      {\Inv \wedge p \in \Nodes \wedge n \notin \Nodes \\ \llbracket add(n, p) \rrbracket}
      {\Inv \wedge n \in \Nodes \wedge n \rightarrow p}
    \end{align*}
  \end{small}
\end{minipage}%
\begin{minipage}{0.45\textwidth}
  \textbf{Remove operation:}
  \begin{small}
    \begin{align*}
      \inferrule[(Remove-Operation)]
      {\Inv \wedge n \neq \troot \\ \llbracket remove(n) \rrbracket}
      {\Inv \wedge n \in \TS}
    \end{align*}
  \end{small}
\end{minipage}

\textbf{Definitions:}
\vspace{-20pt}
\begin{small}
  \begin{align*}
  operation &\triangleq (type, params, priority)\\
  \mathtt{C} &\triangleq set\ of\ concurrent\ operations\\
  \mathtt{H} &\triangleq history\ of\ operations\ available\ at\ the\ origin\ replica\\
  \mathit{crit\mhyphen anc\mhyphen overlap}(op_1, op_2) &\triangleq op_1.params.n \in \mathit{critical\_ancestor}(op_2) \ \wedge \\
  & \quad  \ op_2.params.n \in \mathit{critical\_ancestor}(op_1)\\
  \mathit{self\mhyphen or\mhyphen under}(n) &\triangleq \{n' | n' = n \vee (n' \in \Nodes \wedge n' \rightarrow^* n)\}
\end{align*}
\end{small}
\textbf{Move operation:} 
\begin{small}
  \begin{align*}
    \inferrule[(Up-move-Operation)]
    {\Inv \wedge n \in \Nodes \wedge n \neq \troot \wedge p' \in \Nodes  \wedge\,  n \neq p'\wedge p' \not\rightarrow^* n \wedge \rank(n) > \rank(p')  \\ \llbracket up\mhyphen move(n, p') \rrbracket}
    {{\begin{minipage}[b]{30em} $\echanged{\nexists op \in \mathtt{C} \centerdot op.type = up\mhyphen move \wedge\, op.params.n = n \wedge\, op.priority > priority}$ \\
      $\enew{\nexists op \in \mathtt{H} \centerdot (op.type = up\mhyphen move \wedge\ p' \in \mathit{self\mhyphen or\mhyphen under}(op.params.n))}$ \\
       $\enew{\qquad \vee (op.type = down\mhyphen move \wedge\ (n \in \mathit{self\mhyphen or\mhyphen under}(op.params.n)}$\\
       $\enew{ \qquad \qquad \vee\ (op.params.n \in \mathit{self\mhyphen or\mhyphen under}(n)}$\\ 
       $\enew{\qquad\qquad\qquad\wedge\ p' \in \mathit{self\mhyphen or\mhyphen under}(op.params.n))))}$
    \end{minipage}} \implies  \Inv \wedge n \rightarrow p'}
\end{align*}
\end{small}
\vspace{-10pt}
\begin{small}
  \begin{align*}
    \inferrule[(Down-move-Operation)]
    {\Inv \wedge n \in \Nodes \wedge n \neq \troot \wedge p' \in \Nodes  \wedge\,  n \neq p'\wedge p' \not\rightarrow^* n \wedge \rank(n) \leq \rank(p') \\ \llbracket down\mhyphen move(n, p') \rrbracket}
    {{\begin{minipage}[b]{30em} $\echanged{\nexists op \in \mathtt{C} \centerdot op.type = up\mhyphen move}\\ \echanged{\qquad \wedge\, (crit\mhyphen anc\mhyphen overlap(down\mhyphen move(n, p'), op) \vee\ op.params.n = n)} \\ \echanged{\wedge\, \nexists op \in \mathtt{C} \centerdot op.type = down\mhyphen move} \\ \echanged{\qquad \wedge\, (crit\mhyphen anc\mhyphen overlap(down\mhyphen move(n, p'), op) \vee\ op.params.n = n) } \\ \echanged{\qquad \wedge\, op.priority > priority}$\\
      $\enew{\nexists op \in \mathtt{H} \centerdot (op.type = up\mhyphen move \wedge\ (n \in \mathit{self\mhyphen or\mhyphen under}(op.params.n)}$\\
      $\enew{ \qquad \qquad \vee\ (op.params.n \in \mathit{self\mhyphen or\mhyphen under}(n)}$\\ 
      $\enew{\qquad\qquad\qquad\wedge\ p' \in \mathit{self\mhyphen or\mhyphen under}(op.params.n))))}$ \\
      $\enew{\qquad \vee (op.type = down\mhyphen move \wedge\ p' \in \mathit{self\mhyphen or\mhyphen under}(op.params.n))}$\end{minipage}} \\ \implies \Inv \wedge n \rightarrow p'}
\end{align*}
\end{small}

%% file: sections/evaluation.tex
\section{Evaluation}
\label{sec:evaluation}

\begin{figure*}
    \subfloat[Response time for different conflict rates]{\includegraphics[scale=0.75]{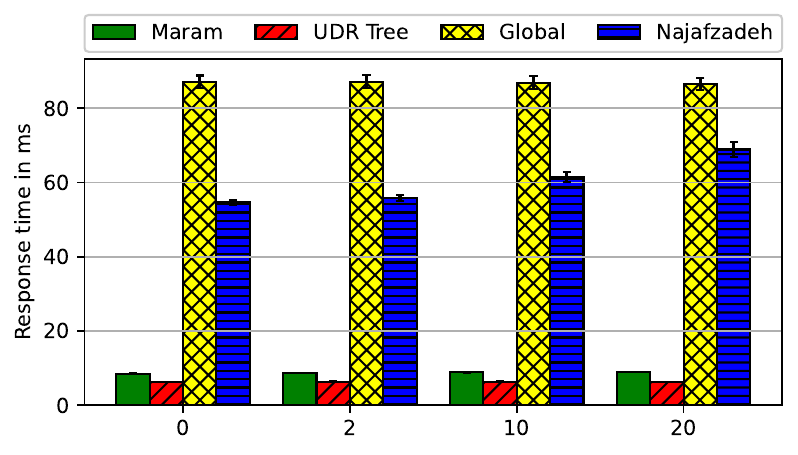}\label{fig:response}}
    \qquad
    \subfloat[Overhead of conflict resolution for different latencies]{\includegraphics[scale=0.75]{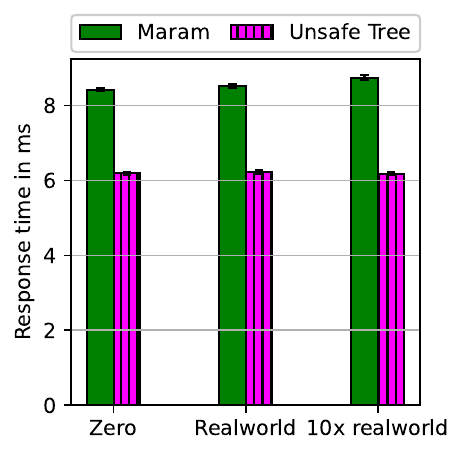}\label{fig:overhead}}
  \qquad
  \subfloat[Stabilization time for different latencies in logarithmic scale]{\includegraphics[scale=0.75]{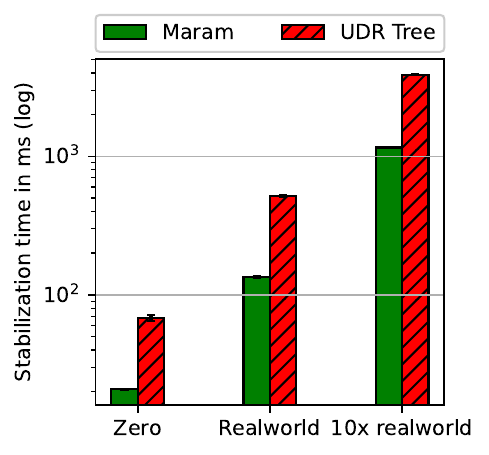}\label{fig:stabilization_log}}
  \label{fig:graphs}
  \caption{Experimental results. Each bar is the average of 15 runs, the error bars show standard deviation}
\end{figure*}

This paper presents the design of a coordination-free, safe, convergent,  and highly available replicated tree.
The specification of \RDTName\ doesn't require any synchronization to execute an operation; this implies that the design is coordination-free.
Sections~\ref{sec:seqdesign} and \ref{sec:refinedspec}  provide a mechanized proof that our design is safe and convergent.
In this section, we conduct an evaluation to showcase the high availability of our design.

We measure availability in two parts - \emph{response time} and \emph{stabilization time}.
The first metric, \emph{response time}, is the time taken to log and acknowledge a client request.
Recall that the effect of a move operation in our specification consists of either updating the state, or a skip.
The effect of the update will be definitive only after being aware of all its concurrent operations. 
In order to measure this, we introduce a metric called \emph{stabilization time}.
Stabilization time measures the duration for which an update is in a transient state.

\renewcommand{\arraystretch}{1.0}
\begin{table}[t!]
 \centering
  \begin{tabular}{@{\extracolsep{0pt}} c >{\raggedleft\arraybackslash}p{1.5cm}  >{\raggedleft\arraybackslash}p{1.5cm}  >{\raggedleft\arraybackslash}p{1.5cm}@{}}
    \toprule
	\multirow{2}{*}{\textbf{Latency}} & \multicolumn{3}{c}{Replicas}  \\
	\cline{2-4}
    & Paris & Bangalore & New York \\
    \hline
     Paris & 0 & 144 & 75 \\
     Bangalore & 144 & 0 & 215 \\
     New York & 75 & 215 & 0 \\
    \bottomrule
  \end{tabular}
  \caption{Real world latency configurations in ms }
  \label{tab:latencyconfig}
\end{table}
\renewcommand{\arraystretch}{1}


We run the experiments\footnote{On DELL PowerEdge R410 machine with 64 GB RAM, and 24 cores @2.40GHz Intel Xeon E5645 processor.}
with three replicas connected in a mesh with a FIFO connection and simulate different network latencies, zero latency, real world latency as shown in \Cref{tab:latencyconfig} and 10 times real world latency.
%
Our warm-up workload, a mix of add, remove and move operations, creates a tree with 997 nodes including the root.
We then have concurrent workloads\footnote{250 operations per replica - 60\% add, 12\% remove, 14\% upmove and 14\% downmove.} on the three replicas, varying conflict rates at 0\%, 2\%, 10\%, and 20\%.  



We compare \RDTName{} with three solutions from the literature: 
\begin{enumerate*}[label=(\roman*)]
\item UDR tree (short for Undo-Do-Redo tree) \cite{DBLP:journals/corr/abs-1805-04263};
\item all move operations acquiring a global lock (Global\_l); and,
\item move operations acquiring read locks on critical ancestors and write lock on the moving node (Subtree\_l) \cite{formel:rep:syn:sh197}.
\end{enumerate*}

The average response time for each design for different conflict rates with latency configuration 2 (Table~\ref{tab:latencyconfig}) are shown in Figure~\ref{fig:response}.
Observe that \RDTName{} and UDR tree show similar average response time across different conflict rates; owing to the synchronization-free design. 
\RDTName\ has a slightly higher response time than UDR tree due to the overhead of metadata calculation.
The response time for Subtree\_l \cite{formel:rep:syn:sh197} increases with an increase in the conflict rate due to lock contention, whereas that of Global\_l is the same across all conflict rates since the proportion of lock-acquiring-moves remains the same.


\Cref{fig:stabilization_log} shows the average stabilization time for our design and the UDR tree design \cite{DBLP:journals/corr/abs-1805-04263} on a logarithmic scale, for different latency configurations.
Our solution gives lower stabilization time, since only down-moves and a very few up-moves have transient state in the case of \RDTName\ whereas  for a UDR tree \cite{DBLP:journals/corr/abs-1805-04263} all operations are in transient state until a local replica asserts that there are no more concurrent operations.%
\footnote{
  Note here that \RDTName{}'s stabilisation time does not depend on conflict
  rate, but only on the proportion of moves in the workload.
  As this proportion grows towards 100\%, the stabilisation time of
  \RDTName{} tends to be the same as that of UDR.
}
%

Next, we run an experiment to measure the overhead introduced by the conflict resolution policy.
As a lower bound, we compare the response time of \RDTName{} with a
na{\"\i}ve unsafe implementation, that uses a simple eventual
consistency approach, and thus is not safe.
\Cref{fig:overhead} shows the response time of both the designs.
For \RDTName\ metadata calculation implies slightly higher response time.
%

%% file: sections/literature.tex
\section{Related work}
\label{sec:literature}
Several works have addressed the problem of designing a replicated tree.
\citet{martin:hal-00648106} introduce some designs for conflict-free replicated tree data types. 
They use set CRDTs to construct replicated trees with different semantics.
Add and remove operations are supported in their design.
However they do not consider move operations.


\citet{DBLP:journals/corr/abs-1805-04263} propose the UDR tree, which supports atomic move operations, using the notion of opsets.
Opsets totally order all operations eventually.
This is more expensive than our solution based on partial order.
When a new operation is performed, all the later operations are redone.
Thus all operations pay a heavy price, and not just the conflicting moves.
But UDR requires eventually consistent delivery layer, whereas \RDTName\ requires a more expensive causal delivery layer.

Compared to the work of \citet{DBLP:journals/corr/abs-1805-04263}, we also show under what conditions a move operation might skip.


\citet{formel:rep:syn:sh197} designs the replicated tree called Subtree\_l in Section~\ref{sec:evaluation}.
Their solution introduces coordination; acquires read locks on the critical ancestors, and a write lock on the node being moved.
This approach is not available under partition, but only move operations pay an overhead.


\citet{fic:rep:sh172} propose a replicated tree with a move operation that does not require any coordination between replicas, replacing each move with non-atomic copy and delete operations.
This might lead to having multiple copies of the same node.

Compared to all the above solutions, our design supports atomic move operation that depends on partial ordering without acquiring any locks.
An atomic update provides all or no guarantee, i.e., either the update is applied or it is not.
Ensuring atomicity avoids partial execution of updates.

\citet{mrdt} introduce the concept of Mergeable Replicated Data Types (MRDTs) inspired by three-way-merge. 
The safety of an MRDT binary tree depends on the labeling of the child-parent relations (whether it belongs to the right or left of the ancestor). 
It also requires to keep track of all the ancestor relations apart from the parent-child relations.
A generic MRDT tree can be considered as an extension to the MRDT binary tree, but requires tracking all ancestor relations and a complex lexicographical ordering when concretizing the merged result. 

%% file: sections/discussion.tex
\section{Discussion}
\label{sec:discussion}

\subsection{Moving from causal consistency to eventual consistency}

\renewcommand{\arraystretch}{1.4}
\begin{table}[t]
 \begin{centering}
  \begin{tabular}{@{\extracolsep{3pt}}c c >{\centering\arraybackslash} p{3cm}  >{\centering\arraybackslash} p{2cm}  >{\centering\arraybackslash} p{5.3cm} @{}}
	\toprule
    	\multicolumn{2}{c}{\multirow{2}{*}{\textbf{Independent}}} &  \multicolumn{3}{c}{Under}\\[-1pt]
	\cline{3-5}
	& & $\mathit{add}(n_2, p_2)$ & $\mathit{remove}(n_2)$ & $\mathit{move}(n_2, p_2')$\\
     	\hline
    	\multirow{4}{*}{\rotatebox[origin=c]{90}{\hspace{0.42cm}Operation}}
		& $\mathit{add}(n_1, p_1)$ 	&  \mbox{\tabfocus{$p_1 \neq n_2$}}
								&  $\tabfocus{true}$ 
								& $\tabfocus{true}$ \\
	\cline{3-5}
	 	& $\mathit{remove}(n_1)$	&  \mbox{\tabfocus{$n_1 \neq n_2$}}
									& $\tabfocus{true}$
									&  $\tabfocus{true}$ \\
	\cline{3-5}
	 &  $\mathit{move}(n_1, p_1')$
	 						&  \mbox{\tabfocus{$n_1 \neq n_2  \vee \newline p_1' \neq n_2$}}
							&  $\tabfocus{true}$
							& \mbox{\tabfocus{ $n_2 \! \notin \! self\mhyphen or\mhyphen under(n_1)$}} \mbox{\tabfocus{$\vee p_1' \notin self\mhyphen or\mhyphen under(n_2)$}}\\
   	\bottomrule
  \end{tabular}
  \caption{Result of dependency analysis. The cell shows the condition under which the operation in the row is independent of the operation in the column.}
  \label{tab:dep_analysis}
 \end{centering}
\end{table}
\renewcommand{\arraystretch}{1}
\citet{Houshmand:2019:HRC:3302515.3290387} propose dependency analysis to help relax the requirement of causal delivery.
We run this analysis for all operations, irrespective of whether the update is tentative or definitive.

\Cref{tab:dep_analysis} shows the results of the dependency analysis of \RDTName{}.
We can observe that no operations are dependent on remove, and add and remove are not dependent on move.
As there is no fully independent operation, relaxing causal delivery is not helpful to \RDTName.

\subsection{Message overhead for conflict resolution}
%
In order to use \RDTName\ in a real-world application, we need to understand the overhead of conflict resolution.
Conflict resolution requires some meta information that is sent along with the update message from the origin replica.
This may have an impact on the bandwidth lost, hence understanding the components is important.

The conflict resolution policy of \RDTName\ needs information to compute a set of concurrent operations.
Assuming a replica works as a single threaded process, we use vector clocks.
The size of vector clocks is linear with the number of replicas.
This poses an additional overhead.

Conflict resolution also takes as input the set of critical ancestors, descendants, and the priority.%
The size of the set of critical ancestors depends on the depth of the subtree comprising the least common ancestor of the node being moved and the destination parent.
The size of the set of critical ancestors is linear to the difference in the rank of the new parent and the least common ancestor.
The size of the set of descendants might be large for the nodes nearer to the root. This poses an overhead on the metadata.
The priority can be a single number or a string and is independent of other factors.
Hence using the conflict resolution of \RDTName\ will cause a considerable overhead on message delivery.

The time taken to compute this metadata is the difference between the response time of a na\"ive unsafe replicated tree and \RDTName\ in \Cref{fig:overhead}.

\subsection{Computing the set of concurrent moves}

\RDTName\ requires a set of concurrent operations to apply the conflict resolution.
For this, the \RDTName\ system layer does not busy-wait.  
Every replica makes progress locally, without waiting to receive remote logs (availability under partition). 
Conflict resolution applies only after a replica receives a concurrent conflicting operation.


To conclude, \RDTName\ is a safe, coordination-free replicated tree, designed using conflict resolution policies.

%% file: sections/conclusion.tex

\section{Conclusion}
This paper presents the design of a light-weight, coordination-free, safe, convergent and highly available replicated tree data structure, \RDTName. 
We provide mechanized proof of safety and convergence of \RDTName, and experimentally demonstrate the efficiency of the design by comparing it with the existing solutions.

%% file: sections/why3spec.tex
\section{Specification of sequential tree in Why3}
\label{app:seqspec}
This section presents the Why3 specification of a sequentially safe tree.
\begin{why3}
use export int.Int
use import map.Map as M
use import set.Fset as F
use seq.Seq, seq.Mem, seq.Distinct

(* auxiliary lemmas on sequences *)
lemma append_empty: forall s: seq 'a.
  s ++ empty == s

lemma empty_length: forall s: seq 'a.
  length s = 0 <-> s == empty
  
predicate disjoint_seq (s1 s2: seq 'a) =
  forall i j. 0 <= i < length s1 ->
		0 <= j < length s2 -> s1[i] <> s2[j]

(* Arbitrary type for a tree node *)
type elt

(* Verifies if two nodes are equal *)
val equal (e1 e2 : elt) : bool
  ensures { result <-> e1 = e2 }

(* Indicates if two nodes are connected by an edge in the tree *)
predicate edge (x y : elt) (f : elt -> elt) =
  x <> y /\ f x = y

(* recursive predicate for expressing a path between two nodes *)
(* in the main text a few cosmetic changes were done, namely *)
(* rename of -f- to -parent- and expand the edge definition *)
predicate path (f: elt -> elt) (x y: elt) (p: seq elt) =
  let n = length p in
  n = 0 /\ x = y
  \/
  n > 0 /\
  p[0] = x /\
  edge p[n - 1] y f /\
  distinct p /\
  (forall i. 0 <= i < n - 1 -> edge p[i] p[i + 1] f) /\
  (forall i. 0 <= i < n     -> p[i] <> y)

predicate reachability (f: elt -> elt) (x y: elt) =
  exists p. path f x y p

(* If there is an edge between nodes x and y
(* the path is defined as the singleton seq with node x  *)
axiom path_to_parent: forall x y : elt, f : elt -> elt.
  edge x y f -> path f x y (cons x empty)
 
(* If there is a path from [from] to [middle] and a path from [middle] to 
  [until] then there is a path from [from] to [until] *)
axiom path_transitivity: forall from middle until f pth1 pth2.
  path f from middle pth1 -> path f middle until pth2 ->
  disjoint_seq pth1 pth2 -> from <> until ->
  (forall j. 0 <= j < length pth1 -> pth1[j] <> until) ->
	path f from until (pth1 ++ pth2)
	
(* Recursive path composition *)
axiom path_composition: forall n x y: elt, f : elt -> elt, pth : seq elt.
  n <> y -> not (mem y pth) ->
  distinct (snoc pth x) -> path f n x pth -> edge x y f ->
  path f n y (snoc pth x)
  
 (* If there is a path between two nodes, that path is unique *)
axiom path_uniqueness: forall x y: elt, f: elt -> elt, 
			      pth1 pth2: seq elt.
  path f x y pth1 -> path f x y pth2 -> pth1 == pth2
  
 (* If node np is not reachable to node c, then np will
    not belong to any path that contains node c *)
axiom path_exclusion: forall f x c np p.
  not (reachability f np c) -> path f x c p -> not (mem np p)

 (* Given a path between two nodes, there is no overlap between any two consecutive subpaths *)
axiom path_separation: forall final initial middle : elt, f : elt -> elt,
		       	      p1 p2 : seq elt.
  path f middle final p2 -> path f initial middle p1 ->
  final <> initial -> middle <> initial -> middle <> final ->
  disjoint_seq p1 p2

constant n: elt (* constant used for defining a state witness *)

type state [@state] = {
  (* parent relation: up-pointers to direct ancestor *)
  mutable parent : elt -> elt;
  (* parent root *)
  mutable root 	 : elt;
  (* nodes in the parent *)
  mutable nodes  : fset elt;
} invariant { F.mem root nodes }
  invariant { parent root = root }
  invariant { forall x. F.mem x nodes -> F.mem (parent x) nodes }
  invariant { forall x. F.mem x nodes -> reachability parent x root }
  invariant { forall x. F.mem x nodes -> 
  		reachability parent root x -> x = root }
  by { parent = (fun _ -> n); root = n; nodes = F.singleton n }

(* Paths already present in the tree remain in the tree after executing 
   the add operation *)
axiom remaining_nodes_add: forall n w p: elt, s: state, l: seq elt.
  path s.parent w s.root l -> not (mem n l) ->
  F.mem w s.nodes -> F.mem p s.nodes -> not (F.mem n s.nodes) ->
  w <> n -> n <> p -> path (M.set s.parent n p) w s.root l

(* Descendants of the node being moved continue to be its descendants *)
axiom descendants_move: forall x c np: elt, f: elt -> elt, p: seq elt.
  x <> np -> c <> np -> x <> c -> not (reachability f np c) ->
  path f x c p -> distinct (cons c p) -> not (mem np p) ->
  (path (M.set f c np) x c p)

(*  Paths nodes unreachable to the node being moved are not affected *)
axiom remaining_nodes_move: forall x c np: elt, s: state, p: seq elt.
  c <> np -> x <> c -> not (reachability s.parent np c) ->
  path s.parent x s.root p -> (not reachability s.parent x c) ->
  distinct p -> (path (M.set s.parent c np) x s.root p)
  
let ghost add (n p : elt) (s : state) : unit
  requires { [@expl:pre_add1] not F.mem n s.nodes }
  requires { [@expl:pre_add2] F.mem p s.nodes }
  ensures  { s.parent = M.set (old s.parent) n p }
  ensures  { edge n p s.parent }
  ensures  { s.nodes = F.add n (old s).nodes }
= s.parent     <- M.set s.parent n p;
  s.nodes      <- F.add n s.nodes;

let ghost remove (n : elt) (s : state) : unit
  requires { [@expl:pre_remove1] forall x. s.parent x <> n }
  requires { [@expl:pre_remove2] n <> s.root }
  ensures  { s.nodes = F.remove n (old s).nodes }
= s.nodes <- F.remove n s.nodes;

let ghost move (c np : elt) (s : state) : unit
  requires { [@expl:pre_move1] F.mem np s.nodes }
  requires { [@expl:pre_move2] F.mem c s.nodes}
  requires { [@expl:pre_move3] not (reachability s.parent np c) }
  requires { [@expl:pre_move4] c <> s.root }
  requires { [@expl:pre_move5] c <> np }
  ensures  { edge c np s.parent }
  ensures  { s.parent = M.set (old s.parent) c np }
= s.parent <- M.set s.parent c np;
\end{why3}

\newpage
\section{CISE analysis on sequential specification}
\label{app:cise3conflicts}

\begin{why3}
use export why3.BuiltIn.BuiltIn
use export why3.Bool.Bool
use export why3.Unit.Unit
use export file_system_alternative.S
use export why3.Tuple2.Tuple2
use export int.Int
use import map.Map as M
use import set.Fset as F
use seq.Seq, seq.Mem, seq.Distinct

predicate same_ext (m1 m2: 'a -> 'b) = forall x: 'a. m1 x = m2 x

val equal_elt (e1 e2 : elt) : bool
 ensures {result <-> e1 = e2}

let ghost predicate state_equality (s1 s2 : state)
 = 
   same_ext s1.parent s2.parent &&
   equal_elt s1.root s2.root &&
   F.(==) s1.nodes s2.nodes

let ghost move_move_analysis (ghost _:()) : (state, state)
   ensures  { match result with
		| x1, x2 -> state_equality x1 x2
		end }  = 
   let ghost c1 = any elt in
   let ghost np1 = any elt in
   let ghost state1 = any state in
   let ghost c2 = any elt in
   let ghost np2 = any elt in
   let ghost state2 = any state in
   assume { (F.mem np1 state1.nodes) /\
	    (F.mem np2 state2.nodes) /\
	    (F.mem c1 state1.nodes) /\ 
	    (F.mem c2 state2.nodes) /\ 
	    (not (reachability state1.parent np1 c1)) /\
	    (not (reachability state2.parent np2 c2)) /\
	    (c1 <> state1.root) /\
	    (c2 <> state2.root) /\
	    (c1 <> np1) /\
	    (c2 <> np2) /\
	    state_equality state1 state2 };
  move c1 np1 state1;
  move c2 np2 state1;
  move c2 np2 state2;
  move c1 np1 state2;
  (state1, state2)

let ghost remove_remove_analysis (ghost _:()) : (state, state)
   ensures  { match result with
		| x1, x2 -> state_equality x1 x2
		end } =
   let ghost n1 = any elt in
   let ghost state1 = any state in
   let ghost n2 = any elt in
   let ghost state2 = any state in
   assume { (F.mem n1 state1.nodes) /\
	    (F.mem n2 state2.nodes) /\
	    (forall x. state1.parent x <> n1 ) /\
	    (forall x. state2.parent x <> n2 ) /\
	    (n1 <> state1.root) /\
	    (n2 <> state2.root) /\
	   state_equality state1 state2 };
   remove n2 state1;
   remove n1 state1;
   remove n1 state2;
   remove n2 state2;
   (state1, state2)

let ghost add_add_analysis (ghost _:()) : (state, state)
   ensures  { match result with
		| x1, x2 -> state_equality x1 x2
		end } =
   let ghost n1 = any elt in
   let ghost p1 = any elt in
   let ghost state1 = any state in
   let ghost n2 = any elt in
   let ghost p2 = any elt in
   let ghost state2 = any state in
   assume { ((not F.mem n1 (nodes state1) /\ F.mem p1 (nodes state1)) /\
	    not F.mem n2 (nodes state2) /\ F.mem p2 (nodes state2)) /\
	    state_equality state1 state2 };
   add n1 p1 state1;
   add n2 p2 state1;
   add n2 p2 state2;
   add n1 p1 state2;
   (state1, state2)

let ghost remove_move_analysis (ghost _:()) : (state, state)
   ensures  { match result with
		| x1, x2 -> state_equality x1 x2
		end } =
   let ghost n1 = any elt in
   let ghost state1 = any state in
   let ghost c2 = any elt in
   let ghost np2 = any elt in
   let ghost state2 = any state in
   assume { (forall x. state1.parent x <> n1 ) /\
	    (n1 <> state1.root) /\
	    (F.mem np2 state2.nodes) /\
	    (F.mem c2 state2.nodes) /\ 
	    (not (reachability state2.parent np2 c2)) /\
	    (c2 <> state2.root) /\
	    (c2 <> np2) /\
	    state_equality state1 state2 };
   move c2 np2 state1;
   remove n1state1;
   remove n1 state2; 
   move c2 np2 state2;
   (state1, state2)

let ghost add_move_analysis (ghost _:()) : (state, state)
   ensures  { match result with
		| x1, x2 -> state_equality x1 x2
		end } =
   let ghost n1 = any elt in
   let ghost p1 = any elt in
   let ghost state1 = any state in
   let ghost c2 = any elt in
   let ghost np2 = any elt in
   let ghost state2 = any state in
   assume { (not F.mem n1 (nodes state1) /\ F.mem p1 (nodes state1)) /\
	    (F.mem np2 state2.nodes) /\
	    (F.mem c2 state2.nodes) /\ 
	    (not (reachability state2.parent np2 c2)) /\
	    (not (reachability state1.parent np2 c2)) /\
	    (c2 <> state2.root) /\
	     (c2 <> np2) /\
	    state_equality state1 state2 };
   add n1 p1 state1;
   move c2 np2 state1;
   move c2 np2 state2;
   add n1 p1 state2;
   (state1, state2)

let ghost add_remove_analysis (ghost _:()) : (state, state)
   ensures  { match result with
		| x1, x2 -> state_equality x1 x2
		end } =
   let ghost n1 = any elt in
   let ghost p1 = any elt in
   let ghost state1 = any state in
   let ghost n2 = any elt in
   let ghost state2 = any state in
   assume { (not F.mem n1 (nodes state1) /\ F.mem p1 (nodes state1)) /\
	    (forall x. state2.parent x <> n2 ) /\
	    (n2 <> state2.root) /\
	    state_equality state1 state2 };
   remove n2 state1;
   add n1 p1 state1;
   add n1 p1 state2;
   remove n2 state2;
   (state1, state2)
\end{why3}

\input{sections/CiseAnalysis.tex}

\section{Specification of \RDTName\ in Why3}
\label{app:concspec}
This section presents the Why3 specification of \RDTName.

\begin{why3}
type state = {
  (* parent relation: up-pointers to direct ancestor *)
  mutable parent     : elt -> elt;
  (* parent root *)
  mutable root 	 : elt;
  (* nodes in the parent *)
  mutable nodes      : fset elt;
  (* rank for each node *)
  mutable rank       : elt -> int;
  (* tombstone nodes  *)
  mutable tombstones : fset elt;
  (* ancestors relation: all ancestors of a node *)
  mutable ancestors  : elt -> fset elt;
} invariant { F.mem root nodes }
  invariant { parent root = root }
  invariant { forall x. F.mem x nodes -> F.mem (parent x) nodes }
  invariant { forall x. F.mem x nodes -> reachability parent x root }
  invariant { forall x. F.mem x nodes ->
		reachability parent root x -> x = root }
  invariant { forall x y. F.mem x nodes /\ F.mem y nodes /\ x <> root /\
		x <> y /\ F.mem y (ancestors x) -> rank x > rank y }
  invariant { forall x. F.mem x nodes /\ x <> root ->
		ancestors x = F.add (parent x) (ancestors (parent x))}
by { parent = (fun _ -> n); root = n; nodes = F.singleton n ; 
     rank = (fun _ -> 1); tombstones = F.empty; 
     ancestors = (fun _ -> F.empty) }

let ghost predicate state_equality (s1 s2 : state)
= same_ext s1.parent s2.parent /\
  equal_elt s1.root s2.root /\
  F.(==) s1.nodes s2.nodes ) /\
  same_ext s1.rank s2.rank /\
  same_ext s1.ancestors s2.ancestors /\
  F.(==) s1.tombstones s2.tombstones

val ghost add (n p : elt) (s : state) : unit
  requires { [@expl:add1] not F.mem n s.nodes }
  requires { [@expl:add2] F.mem p s.nodes }
  writes   { s.parent, s.nodes, s.rank, s.ancestors }
  ensures  { s.parent = M.set (old s.parent) n p }
  ensures  { edge n p s.parent }
  ensures  { s.nodes = F.add n (old s).nodes }
  ensures  { s.rank = M.set (old s).rank n ((s.rank p) + 1) }
  ensures  { s.ancestors = M.set ((old s).ancestors) n
		(F.add p ((old s).ancestors p)) }

val ghost remove (n : elt) (s : state) : unit
  requires { n <> s.root }
  writes   { s.tombstones }
  ensures  { s.tombstones = F.add n (old s).tombstones }

val ghost move (c np : elt) (s : state) : unit
  requires { F.mem np s.nodes }
  requires { F.mem c s.nodes }
  requires { not (reachability s.parent np c) }
  requires { c <> s.root }
  requires { c <> np }
  writes   { s.parent, s.ancestors, s.rank }
  ensures  { [@expl:post1] s.parent = M.set (old s.parent) c np }

let ghost move_refined (c1 np1 c2 np2 : elt) (pr1 pr2 : int)
		       (s : state) : unit
  requires { pr1 <> pr2 }
  requires { F.mem c1 s.nodes /\ F.mem c2 s.nodes }
  requires { F.mem np1 s.nodes /\ F.mem np2 s.nodes }
  requires { not (reachability s.parent np1 c1) /\ 
  	     not (reachability s.parent np2 c2) }
  requires { c1 <> s.root /\ c2 <> s.root }
  requires { c1 <> np1 /\ c2 <> np2 }
  ensures  { (s.parent = M.set (old s.parent) c2 np2) \/
             (s.parent = M.set (old s.parent) c1 np1) \/
             (same_ext s.parent (old s).parent) }
= if (equal_elt c1 c2) then
  if (pr1 < pr2) then move c2 np2 s else move c1 np1 s
  else if (s.rank np1 < s.rank c1) then move c1 np1 s
  else if (F.mem c2 (diff (F.add np1 (s.ancestors np1))
           s.ancestors c1))) then
  if (s.rank np2 < s.rank c2) then ()
  else if (pr1 < pr2) then move c2 np2 s else move c1 np1 s
  else move c1 np1 s
\end{why3}

%% file: sections/CiseAnalysis.tex

\subsection{Why3 proof sessions}
\label{sec:why3-proof-sessions}

This section presents the quantitative results of using Why3 and CISE3 to
\mbox{analyse} the sequential operations. Each table presents the set of
generated verification conditions for a specific pair of operations, whose names
appear in the head of the table. For each verification condition, we run the
available SMTs until one of them is able to discharge it, or else every one
fails to complete the proof. Finally, each verification condition is identified
with a name of the form \texttt{lemma P}, where~\texttt{P} for the nature of the
condition, \emph{i.e.}, it is either a precondition or a postcondition. For
instance, on the first table, \texttt{pre\_move1} stands for the first
precondition of the \texttt{move} operation, and so on. Times are given in
seconds.

\begin{table}[h]
\begin{tabular}{|l|l|l|l|l|l|l||c|}
	\hline \multicolumn{2}{|c|}{Proof obligations } & \provername{Alt-Ergo 2.3.2} & \provername{CVC4 1.7} & \provername{Z3 4.8.6} \\
	\hline
	\explanation{VC for }  
	& \explanation{pre\_move1} & \noresult& \valid{0.10} & \noresult\\
	\cline{2-5}
	\continuation{move\_move\_analysis}  & \explanation{pre\_move2} & \noresult& \valid{0.10} & \noresult\\
	\cline{2-5}
	& \explanation{pre\_move3} & \noresult& \valid{0.07} & \noresult\\
	\cline{2-5}
	& \explanation{pre\_move4} & \noresult& \valid{0.10} & \noresult\\
	\cline{2-5}
	\cline{2-5}
	& \explanation{pre\_move5} & \noresult& \valid{0.11} & \noresult\\
	\cline{2-5}
	& \explanation{pre\_move1} & \noresult& \valid{0.21} & \noresult\\
	\cline{2-5}
	& \explanation{pre\_move2} & \noresult& \valid{0.22} & \noresult\\
	\cline{2-5}
	& \explanation{pre\_move3} & \timeout{1s} & \timeout{1s} & \timeout{1s} \\
	\cline{2-5}
	& \explanation{pre\_move4} & \noresult& \valid{0.15} & \noresult\\
	\cline{2-5}
	\cline{2-5}
	& \explanation{pre\_move5} & \noresult& \valid{0.09} & \noresult\\
	\cline{2-5}
	& \explanation{pre\_move1} & \noresult& \valid{0.11} & \noresult\\
	\cline{2-5}
	& \explanation{pre\_move2} & \noresult& \valid{0.10} & \noresult\\
	\cline{2-5}
	& \explanation{pre\_move3} & \noresult& \valid{0.09} & \noresult\\
	\cline{2-5}
	& \explanation{pre\_move4} & \noresult& \valid{0.09} & \noresult\\
	\cline{2-5}
	\cline{2-5}
	& \explanation{pre\_move5} & \noresult& \valid{0.11} & \noresult\\
	\cline{2-5}
	& \explanation{pre\_move1} & \noresult& \valid{0.24} & \noresult\\
	\cline{2-5}
	& \explanation{pre\_move2} & \noresult& \valid{0.22} & \noresult\\
	\cline{2-5}
	& \explanation{pre\_move3} & \timeout{1s} & \timeout{1s} & \highfailure \\
	\cline{2-5}
	& \explanation{pre\_move4} & \noresult& \valid{0.14} & \noresult\\
	\cline{2-5}
	\cline{2-5}
	& \explanation{pre\_move5} & \noresult& \valid{0.09} & \noresult\\
	\cline{2-5}
	& \explanation{postcondition}  & \valid{0.26} & \timeout{1s} & {\highfailure} \\
	\hline \end{tabular}
\end{table}

\begin{table}
\begin{tabular}{|l|l|l|l|l|l|l||c|}
	\hline \multicolumn{2}{|c|}{Proof obligations } & \provername{Alt-Ergo 2.3.2} & \provername{CVC4 1.7} & \provername{Z3 4.8.6} \\
	\hline
	\explanation{VC for} & \explanation{pre\_remove1} & \noresult& \valid{0.16} & \noresult\\
	\cline{2-5}
	\continuation{remove\_remove\_analysis} & \explanation{pre\_remove2} & \noresult& \valid{0.22} & \noresult\\
	\cline{2-5}
	& \explanation{pre\_remove1} & \noresult& \valid{0.20} & \noresult\\
	\cline{2-5}
	& \explanation{pre\_remove2} & \noresult& \valid{0.16} & \noresult\\
	\cline{2-5}
	& \explanation{pre\_remove1} & \noresult& \valid{0.27} & \noresult\\
	\cline{2-5}
	& \explanation{pre\_remove2} & \noresult& \valid{0.21} & \noresult\\
	\cline{2-5}
	& \explanation{pre\_remove1} & \noresult& \valid{0.24} & \noresult\\
	\cline{2-5}
	& \explanation{pre\_remove2} & \noresult& \valid{0.15} & \noresult\\
	\cline{2-5}
	& \explanation{postcondition}& \noresult& \valid{0.10}
        & \noresult \\
	\cline{2-5}
	& &  \noresult& \valid{0.09} & \noresult \\
	\cline{2-5}
	& & \noresult& \valid{0.23} & \noresult \\

	\hline
	\explanation{VC for}   & \explanation{pre\_add1} & \noresult& \valid{0.11} & \noresult\\
	\cline{2-5}
	\continuation{add\_add\_analysis}& \explanation{pre\_add2} & \noresult& \valid{0.11} & \noresult\\
	\cline{2-5}
	& \explanation{pre\_add1} & \timeout{1s} & \timeout{1s} & \highfailure \\
	\cline{2-5}
	& \explanation{pre\_add2} & \noresult& \valid{0.22} & \noresult\\
	\cline{2-5}
	& \explanation{pre\_add1} & \noresult& \valid{0.10} & \noresult\\
	\cline{2-5}
	& \explanation{pre\_add2} & \noresult& \valid{0.11} & \noresult\\
	\cline{2-5}
	& \explanation{pre\_add1} & \timeout{1s} & \timeout{1s} & \timeout{1s} \\
	\cline{2-5}
	& \explanation{pre\_add2} & \noresult& \valid{0.22} & \noresult\\
	\cline{2-5}
	& \explanation{postcondition} & \valid{0.51} & \timeout{1s} & \highfailure \\

	\hline
	\explanation{VC for}  
	& \explanation{pre\_move1} & \noresult& \valid{0.24} & \noresult\\
	\cline{2-5}
	\continuation{remove\_move\_analysis} & \explanation{pre\_move2} & \noresult& \valid{0.23} & \noresult\\
	\cline{2-5}
	& \explanation{pre\_move3} & \noresult& \valid{0.93} & \noresult\\
	\cline{2-5}
	& \explanation{pre\_move4} & \noresult& \valid{0.15} & \noresult\\
	\cline{2-5}
	\cline{2-5}
	& \explanation{pre\_move5} & \noresult& \valid{0.11} & \noresult\\
	\cline{2-5}
	& \explanation{pre\_remove1} & \timeout{1s} & \timeout{1s} & \highfailure \\
	\cline{2-5}
	& \explanation{pre\_remove2} & \noresult& \valid{0.23} & \noresult\\
	\cline{2-5}
	& \explanation{pre\_remove1} & \noresult& \valid{0.19} & \noresult\\
	\cline{2-5}
	& \explanation{pre\_remove2} & \noresult& \valid{0.19} & \noresult\\
	\cline{2-5}
	& \explanation{pre\_move1} & \noresult& \valid{0.32} & \noresult\\
	\cline{2-5}
	& \explanation{pre\_move2} & \timeout{1s} & \timeout{1s} & \highfailure \\
	\cline{2-5}
	& \explanation{pre\_move3} & \noresult& \valid{0.15} & \noresult\\
	\cline{2-5}
	& \explanation{pre\_move4} & \noresult& \valid{0.17} & \noresult\\
	\cline{2-5}
	\cline{2-5}
	& \explanation{pre\_move5} & \noresult& \valid{0.11} & \noresult\\
	\cline{2-5}
	& \explanation{postcondition} &
                      \noresult& \valid{0.22} &\noresult \\
	\cline{2-5}
	& & \noresult& \valid{0.16} & \noresult \\
	\cline{2-5}
	& & \noresult& \valid{0.23} & \noresult \\
	\hline
	\end{tabular}
\end{table}

\begin{table}
\begin{tabular}{|l|l|l|l|l|l|l||c|}
	\hline \multicolumn{2}{|c|}{Proof obligations } & \provername{Alt-Ergo 2.3.2} & \provername{CVC4 1.7} & \provername{Z3 4.8.6} \\
	\hline
	\explanation{VC for} & \explanation{pre\_add1} & \noresult& \valid{0.12} & \noresult\\
	\cline{2-5}
	\continuation{add\_move\_analysis}& \explanation{pre\_add2} & \noresult& \valid{0.11} & \noresult\\
	\cline{2-5}
	 & \explanation{pre\_move1} & \noresult& \valid{0.35} & \noresult\\
	\cline{2-5}
	& \explanation{pre\_move2} & \noresult& \valid{0.41} & \noresult\\
	\cline{2-5}
	& \explanation{pre\_move3} & \noresult& \valid{0.11} & \noresult\\
	\cline{2-5}
	& \explanation{pre\_move4} & \noresult& \valid{0.15} & \noresult\\
	\cline{2-5}
	\cline{2-5}
	& \explanation{pre\_move5} & \noresult& \valid{0.09} & \noresult\\
	\cline{2-5}
	& \explanation{pre\_move1} & \noresult& \valid{0.10} & \noresult\\
	\cline{2-5}
	& \explanation{pre\_move2} & \noresult& \valid{0.22} & \noresult\\
	\cline{2-5}
	& \explanation{pre\_move3} & \noresult& \valid{0.12} & \noresult\\
	\cline{2-5}
	& \explanation{pre\_move4} & \noresult& \valid{0.11} & \noresult\\
	\cline{2-5}
	\cline{2-5}
	& \explanation{pre\_move5} & \noresult& \valid{0.11} & \noresult\\
	\cline{2-5}
	& \explanation{pre\_add1} & \noresult& \valid{0.31} & \noresult\\
	\cline{2-5}
	& \explanation{pre\_add2} & \noresult& \valid{0.34} & \noresult\\
	\cline{2-5}
	& \explanation{postcondition}  & \noresult& \valid{0.35} & \noresult\\
	\cline{2-5}
	&  &  \noresult& \valid{0.20} & \noresult\\
	\cline{2-5}
	&  & \noresult& \valid{0.35} & \noresult\\

	\hline
	\explanation{VC for}   & \explanation{pre\_remove1} & \noresult& \valid{0.17} & \noresult\\
	\cline{2-5}
	\continuation{add\_remove\_analysis}& \explanation{pre\_remove2} & \noresult& \valid{0.16} & \noresult\\
	\cline{2-5}
	& \explanation{pre\_add1} & \noresult& \valid{0.19} & \noresult\\
	\cline{2-5}
	& \explanation{pre\_add2} & \timeout{1s} & \timeout{1s} & \highfailure \\
	\cline{2-5}
	& \explanation{pre\_add1} & \noresult& \valid{0.25} & \noresult\\
	\cline{2-5}
	& \explanation{pre\_add2} & \noresult& \valid{0.23} & \noresult\\
	\cline{2-5}
	& \explanation{pre\_remove1} & \noresult& \valid{0.29} & \noresult\\
	\cline{2-5}
	& \explanation{pre\_remove2} & \noresult& \valid{0.11} & \noresult\\
	\cline{2-5}
	& \explanation{postcondition}  & \noresult& \valid{0.27} & \noresult \\
	\cline{2-5}
	& & \noresult& \valid{0.19} & \noresult  \\
	\cline{2-5}
	& & \timeout{1s} & \timeout{1s} & \timeout{1s} \\
	\hline
	\end{tabular}
\end{table}